\documentclass[12pt,preprint]{aastex}

\usepackage{multirow}
\usepackage{graphicx}
\usepackage[caption=false]{subfig}

\begin{document}

\title{ {\it Chandra} Detection of An Evolved Population of Young Stars in Serpens South.}

\shorttitle{{\it Chandra}'s Evolved YSOs in Serpens South.}

\author{E. Winston\altaffilmark{1}, S. J. Wolk\altaffilmark{1}, R. Gutermuth\altaffilmark{2}, T.L. Bourke\altaffilmark{3}} 

\altaffiltext{1}{Harvard Smithsonian Center for Astrophysics, 60 Garden St., Cambridge MA 02138, USA.} 
\email{elaine.winston@cfa.harvard.edu}
\altaffiltext{2}{Department of Astronomy, University of Massachusetts, Amherst, MA 01003 }
\altaffiltext{3}{SKA Organisation, Jodrell Bank Observatory, Cheshire SK11 9DL, UK.}

\begin{abstract}

We present a {\it Chandra} study of the deeply embedded Serpens South star-forming region, examining cluster structure and 
disk properties at the earliest stages.   In total, 152 X-ray sources are detected. Combined with {\it Spitzer} 
and 2MASS photometry, 66 X-ray sources are reliably matched to an IR counterpart. 
We identify 21 class I, 6 flat spectrum,  16 class II, 
and 18 class III young stars; 5 were unclassified. 
Eighteen sources were variable in X-rays: 8 exhibiting flare-like emission, and one periodic source. 
The cluster’s X-ray luminosity distance was estimated: the best match was to the nearer distance of 260~pc for the front of the Aquila Rift complex. 
The $N_{H}$ vs. $A_{K}$ ratio is found to be $\sim$0.68x10$^{22}$, similar to that measured in other young low mass regions, 
but lower than that measured in the ISM and high mass clusters ($\sim$1.6-2x10$^{22}$).   
We find the spatial distribution closely follows that of the dense filament from which the stars have formed, with the class II 
population still strongly associated with the filament.   There are four sub-clusters in the field, with three forming knots in 
the filament, and a fourth to the west, which may not be associated but may be contributing to the distributed class III population.
 A high percentage of diskless class IIIs ( upper limit 30\% of classified X-ray sources) in such a young cluster  could indicate that 
processing of disks is influenced by the cluster environment and is not solely time-scale dependent.

\end{abstract}


\today

\section{\bf Introduction}

The study of clustered star formation has come of age in the last decade with the advent of space based X-ray and infrared/optical missions 
such as {\it Chandra} and Spitzer, which can provide a more complete census of nearby clusters including both disk-bearing and diskless young stars (e.g. \citet{2013ApJS..209...31P}). 
There have been a number studies of nearby young stellar clusters which combine the emission in the high energy X-ray regime with the IR emission of the 
pre-main sequence stars and protostars \citep{wol, win10, 2017ApJS..229...28G}.

The Serpens South cluster is an extremely young and deeply embedded region of low mass star formation. 
it is inferred to be the youngest known such region in the local galaxy from its protostellar/pre-main sequence fraction, making it  a unique test-bed for 
the study of disk evolution in clustered environments.   
It was first identified in Spitzer observations of the Gould Belt regions by \citet{gut08}, and was associated with the Serpens cloud, part of the Aquila Rift. 
The initial Spitzer study of the cluster found it to contain 92 young stellar objects (YSOs), of which half were class II and half class I, giving Serpens South 
one of the highest protostellar fractions known in young stellar clusters. 
It has also been associated with W40 by \citet{kony} due to its apparent close proximity and the cloud structure visible at Herschel wavelengths. 

The assumed distance to Serpens South has thus depended on whether it is considered to 
be at the same distance as W40 of 550-900pc \citep{kuhn, radha, smith}, or the Aquila region at 260-415 pc \citep{str, dzib2010}.  \citet{gut08} compared the 
local standard of rest (LSR) velocities of two protostars, taken with the SMA, with those of Serpens Main \citep{white} and W40 \citep{zhu} and found that 
they are most similar to Serpens Main, believed to be $\sim$260pc at that time. 
\citet{dzib2011} measured the trigonometric parallax to the YSO EC~95 in Serpens Main as 429$\pm$2pc.   
Recent results from the 'GOBELINS' $VLBA$ survey measured the parallax to seven young stars in the W40 and Serpens Main fields and found that all sources 
are similarly distant, with a mean distance to both regions of 436$\pm$9pc \citep{ortiz}.  This would imply that W40 and the Serpens clusters may be part of one large massive 
star-forming complex, the nearest after the Orion complex.   However, to date, there have been no VLBA distance measurements to any young stars associated  
with Serpens South, and given the close proximity of the two regions on the sky, there remains the possibility of contamination from Serpens South in the W40 field.

X-ray observations may provide an independent estimate of the distance to young stellar clusters, while 
allowing us to examine the high-energy properties of the young stars, which are more usually associated with longer wavelengths in the 
infrared where their circumstellar material glows brightly in reprocessed starlight. 
\citet{fei} noted a 'universal X-ray luminosity function' for young stellar clusters ( $<\sim$20~Myrs old) with a normal distribution dependant only on the distance to the cluster; this distribution has been used to estimate distances to a number of nearby regions, e.g. Serpens Main and W~40 \citep{win10, kuhn}.     
The XLF has been used to buttress the distance estimates for LkH$\alpha$~101 \citep{2009ApJ...691.1128O} and to determine the relative distances to regions in Orion \citep{2014AA...564A..29B, 2016ApJ...820L..28P}.
It is therefore a useful tool in making a comparative estimate to the distances of young clusters.   We will employ this technique in the discussion of the 
relation of Serpens South to Serpens Main and W~40.   

The spatial structure of young clusters and the distribution of their YSOs by evolutionary class can provide an insight into the formation history of the region, 
the relative age of any subclusters, and whether any external triggering is likely to have occured.  The association of the stellar cluster with underlying filamentary structure observed 
at far-IR wavelengths can also be inferred.   
\citet{naka} reported the first observation of CCS ($J_N = 4_3 - 3_2$), with a lifetime of $\sim10^5$yr, 
in a cluster-forming region and they identify six ridges which appear to be colliding to form the Serpens South cluster.  
\citet{2013ApJ...766..115K} observed the southern filament associated with Serpens South in N$_2$H$^+$ and found that material was both radially contracting onto the
filament, and accreting along the filaments long axis at a rate sufficient to support the current rate of star formation.    
\citet{freisen}  report $NH_3$ observations showing that the kinematics of the cluster and surrounding filaments are complex with heirarchical structure.  
The low virial parameters observed suggest the magnetic field in the region is not enough to support the filaments against collapse.  

In young stellar clusters the lifetimes of circumstellar disks are thought to be a few millions years \citep{her2} while the estimated age of Serpens South is less than 1 Myrs. 
It is therefore of great importance to determine whether an older population of class III diskless YSOs exists in this young cluster and what proportion of the clusters young 
stars have reached this stage in their evolution.    
Such studies are important to distinguish between the temporal evolution of circumstellar material, where the class III stars are an older stage to the class II, and evolution 
due to the impact of external factors such as competitive accretion, disk and/or envelope stripping from tidal interactions or early ejection of stars from their natal accretion sites \citep{pfalzner}. 
In this latter case, we would expect that the class III and class II members of the cluster would have the same age, as fitted by evolutionary models on the HR diagram.    
These studies may also impact the estimated ages of young clusters determined from their IR disk fractions - if disk frequency is heavily impacted by cluster environment 
then these ages will not be meaningful unless they are quantified by a measure of the interaction frequency of the YSOs within the cluster. 
For this purpose, the X-ray disk frequency has the important benefit of reliably distinguishing the diskless class III stars (which do not exhibit excess emission in the IR) 
from field stars in combination with the IR photometry.

In this paper we first discuss, in Sect.~\ref{obs}, the {\it Chandra} and Spitzer observations and the data reduction process. 
In Sect.~\ref{iding} we will discuss the identification and evolutionary classification of the YSOs. 
A discusssion of the X-ray properties of the identified members, their spatial distribution, and the X-ray luminosity distance determination is presented in Sect.~\ref{disc}. Finally, a brief summary is presented in Sect.~\ref{summ}.

\section{Observations and Data Reduction}\label{obs}

\subsection{{\it Chandra} ACIS-I Data}

The Serpens South cluster was observed by {\it Chandra} with the ACIS instrument on 7th June 2010,  in a single epoch with a 97.49~ks/pix  exposure time as part of OBSID~11013. 
The ACIS-I field was centered on the J2000 coordinates: 18$^{h}$30$^{m}$03$^{s}$, -02$^o$01$'$58$''$,  with a 17'$\times$17' field of view.  
The observation was taken with ACIS-I in $FAINT$ mode, and $TE$ exposure mode with no grating, and included two of the ACIS-S chips adjacent to the ACIS-I array.  

The primary and secondary data were downloaded from the Chandra X-ray Center's archive and reprocessed with the $CIAO$~v4.8 
$reproc$ tool to ensure that the latest CALDB~4.7.2 corrections were applied \citep{ciao}.   The $CIAO$ task $fluximage$ was used to create an 
exposure map to correct for the photon energy dependent effective collecting area and the presence of chip gaps.   
Exposure maps were created to accurately represent the effective area across the imaging array.\footnote{we assumed a median energy of 1.7keV when calculating the 
effective area. }   The effective area is applied by $CIAO$ to all further processing. 
Source detection was performed using $wavdetect$ with scales from 1''-16'' 
The psf enclosing 90\% of the X-ray flux was generated using $mkpsfmap$ and was chosen to reduce the merging of sources in the $wavdetect$ run.   
The $roi$ tool was used to determine source and background regions without overlaps and a background radius from 3 to 6 times the source PSF size.  
The final source catalogue contained 152 X-ray sources in the Serpens South field,  as listed in Table~\ref{IRphot1}. 
The point source luminosity limit was estimated from the \citet{fei} COUP data as $log_{10}~L_{X} \sim 28.4~ ergs~s^{-1}$ assuming a distance of 430~pc, 
or  $log_{10}~L_{X} \sim 28.0~ ergs~s^{-1}$ assuming a distance of 260~pc. 
 The astrometry of the ACIS-I sources has subarcsec ($\sim$0.5'') spatial precision within $\sim$5' of the aim point.

\subsubsection{Chandra Spectral Analysis}
Spectral analysis of each of the 152 sources was performed to obtain a measure of the bulk temperature of the stellar corona and the hydrogen 
column density along the line of sight, using $CIAO$'s $Sherpa$ package \citep{sherpa}.  For sources with more than 30~counts, a one-temperature 
Raymond-Smith plasma model \citep{ray}  combined with an absorption model component was fit to the data, using the Levenberg-Marquardt 
optimization method and $\chi$-squared statistics.   The $covariance$ routine was employed to obtain a measure of the uncertainties associated 
with $N_H$ and $kT$.    The model's intial conditions were set to $N_H = 10^{21}cm^{-2}$ and $kT = 1.0keV$.    
The modeled flux was calculated using $calc\_energy\_flux$ for the range 0.3-8.0keV and the three CXC bands [soft: 0.3-0.9, medium: 0.9-1.5, hard: 1.5-8.0keV].  
Both the absorbed and unabsorbed fluxes were calculated; the unabsorbed flux by using the Raymond-Smith component of the model only.  
The median and uncertainties on the absorbed and unabsorbed model fluxes were measured using $sample\_flux$ and a sample size of 1000 for the four bands.  
For sources with greater than 100~counts, a two-temperature Raymond-Smith fit was also performed.

\subsubsection{Variability}

The variability of the X-ray sources was assessed using the Gregory-Loredo algorithm \citep{gregory}, which tests for periodic signals using 
maximum-likelihood statistics to evaluate a large number of possible break points from the prediction of constancy.  
The $N$ events for each source are binned in histograms with $m$ bins, with $m$ ranging from 2 to $m_{max}$, where $m_{max}$ is set so that time scales 
for variability down to 50s are considered.    The algorithm returns an odds ratio for $m$ bins versus a
flat light curve.   The resultant light curve is constructed by weighting the binnings by their odds ratios.        
The algorithm can be used to detect non-periodic variability, such as flaring, by forcing the period to equal the length of the observation.   
 Full descriptions of the Gregory-Loredo variability algorithm\footnote{http://cxc.harvard.edu/newsletters/news\_13/gl\_algo.html}   
and the $glvary$ routine\footnote{http://cxc.harvard.edu/ciao/ahelp/glvary.html}  are available on the $CIAO$ website.

The $CIAO$ routine $glvary$ was used to search for source variability in the observation:
To begin, the routine $dither\_region$ was used to remove the effects of instrumental dithers moving sources on and off the ACIS chips by calculating the 
fractional area for each source.   The routine $glvary$ was then run using the $roi$ region files and the fractional areas as input to construct lightcurves for 
all 152 sources and a variability index indicating whether the source is statistically likely to have varied during the 97.5~ks or 1.13~day observation. 
The probability that the source is variable is calculated, but is inconclusive for  $P\sim$0.5-0.9, and so a second criterion based on the average count rate 
and average standard deviation of the light curve is used to calculate the returned variability index.  The index ranges from 0-10; sources with values above 
6 are considered to be variable.

\subsection{Near \& Mid-IR data}

Serpens South was observed with {\it Spitzer} \citep{werner} as part of the 5.1 square degree Serpens-Aquila Rift observation in the Gould Belt Legacy Survey, PID~30574, 
on 27th October 2006.   IRAC imaging \citep{faz}  at 3.6, 4.5, 5.8, 8.0$\mu$m was obtained in High Dynamic Range mode, which obtained 0.4 and 10.4 second intergrations 
for four dithered images at each mosaic postion.  
The Basic Calibrated Data (BCD) products were produced by the Spitzer Science Center's pipeline v18.5. The data were reduced and combined using an updated version of the \citet{gut09} 
method based on the IDL $ClusterGrinder$ package described in \citet{gut08}.    Improvements include the treatment of bright source artifacts and the correction of astrometric 
offsets by BCD image ahead of the final mosaic. A refined pixel scale of 0.87''/pix provides improved centroiding and separation of close sources as described in  \citet{gunther} and \citet{masiunas}.
Further improvement to the method, based on that of \citet{mizuno} for MIPSGAL, provides enhanced global background matching.  This improves the overall appearance of the mosaic,  cosmic ray rejection, and reduces the effects of large scale gradients on the estimation of the background flux \citep{gut2018}.   
 MIPS mosaics \citep{rie} (24, 70, 160$\mu$m) were taken at medium scan rate with full-width scan stepping.   
Due to the high background and low angular resolution of the 70 and 160$\mu$m bands, point source extraction was only applied to the 24$\mu$m mosaic.   
The mid-IR photometry was supplemented by $J$, $H$ and $K$-band photometry from the 2MASS point source catalogue \citep{skr}.   
The photometric catalogues were merged using a maximum matching radius of 1''.  The typical radial residual between the 2MASS photometry and each IRAC band was $<$0.1'' rms, 
with $>$99\% of IRAC sources with 0.5'' radial distance of the 2MASS position.     
In matching the 2MASS and MIPS sources, we require a detection in at least one IRAC band prior to inclusion of that source in the catalog.  
The final source position used was the mean of all detected positions, with uniform weights across all bands (and counting the 2MASS positions per 2MASS bandpass detected, although 
those positions are already merged by band from that dataset. 
There were a total of 33,199 sources in our IR catalog covering the Serpens South field.  

A three color image of the Serpens South in context with the W~40 region is shown in Fig~\ref{fig1}.   
The dense filament running through the region is visible as a dark dust lane even at 24$\mu$m.    
The white  diamond shows the limits of the X-ray field; it is clear from the underlying image that there are regions of active star formation, indicated by  24$\mu$m (red) 
sources, just off the northern and southern edges of the X-ray field.  Similarly, there is a site of star formation $\sim$5' off the western edge, indicated by 
the dark spot and the diffuse emission.   We have therefore expanded our IR analysis to include an area beyond the $Chandra$ field (the dotted square in Fig.~\ref{fig1})  in order to account for distributed populations of YSOs which may have come from outside the field of view of the X-ray pointing.

\section{ YSO Identification \& Classification}\label{iding}

Young stellar objects are most frequently identified by their excess emission at IR wavelengths. This emission arises from reprocessed 
stellar radiation in the dusty material of their natal envelopes or circumstellar disks. The infrared identification of YSOs is 
carried out by identifying sources that possess colors indicative of IR excess and distinguishing them from reddened and/or 
cool stars \citep{meg, all, gut1}.

Young stellar objects have also been observed to possess elevated levels of X-ray emission compared to main sequence stars. 
With luminosities, $L_X$,  $\sim10^{3.5}$ times those of their main sequence counterparts, we can distinguish them from 
foreground or background field stars \citep{fei2, fei3}. By taking advantage of this 
property we can identify YSOs that do not have emission from a dusty disk (evolutionary class III) and would otherwise be 
indistinguishable from the field stars. Both protostars (class 0/I/FS) and pre-main sequence stars with disks (class II/transition disks) 
may also be detected via their elevated X-ray emission.

Table~\ref{sid}  lists the number of YSOs detected by class in the IR-field, the Chandra field of view, and the X-ray detected sample.
Table~\ref{IRphot1} list the source IDs, positions, and properties of the detected X-ray sources.
Table~\ref{IRphot} lists the IR photometry of the counterparts to X-ray sources; the calculated extinction at K-band and evolutionary classification is included.   
Table~\ref{IRphotYSOs} lists the IR photometry for all YSOs identified by our improved \citet{gut09} method in the SS IR-field.  
In Sec.~\ref{irysos} we will first discuss our methodology for selecting cluster members in the IR, before going on to detail the selection and classification 
of the X-ray luminous YSOs based on their IR properties in Sec.~\ref{xrayysos}.

\subsection{IR-excess YSOs}\label{irysos}

We define a $\sim$0.6 square degree field  (henceforth the Serpens South IR-field (SSIR) ) centred on the Serpens South {\it Chandra} pointing, which covers 
the extended Serpens South filament identified in the {\it Herschel} column density maps of \citet{kony} and extending just beyond the corners of the 
{\it Chandra} field of view, as shown in Fig.~\ref{fig1}.  The W~40 region is excluded from this field, though it is possible that some sources are interlopers from that region.  

The YSOs in the SS IR-field were classified by \citet{gut2018}, following their methods described in \citet{gut08} and \citet{gut09} with two improvements:
We increase from 1-$\sigma$ to 3-$\sigma$ the uncertainty requirement in $K-[3.6]$v$[3.6-4.5]$ color space to remove field star contaminants from the YSO selection.  
We no longer require transition disks to have $[3.6]<14$ to improve identification in more distant regions where they risked being misclassified as embedded objects. 
This approach removed likely contaminants from the catalog:  PAH galaxies, AGN, and knots of emission which may be mistaken for YSOs.
Photometric uncertainties $< 0.2$~mags were required in all bands used for a {\it particular} color-color diagram.  
Sources with colors and magntiudes consistent with YSOs were classified as class 0 / class 1: 
for (deeply) embedded protostars,  class 2: disk-bearing pre-main sequence objects, and class 3: objects with weak/anaemic disks.  
This classification, $class_{color}$, uses arabic numerals to distinguish it from the SED-slope based classification, $class_{SED}$, of the IR-detected X-ray sources.  

There were 33,199 sources in the SSIR-field, of which 299 were identified as YSOs based on the $class_{color}$ classification method.  These were classified as follows: 90 class 0/1, 195 class 2 sources, and 14 were anaemic disk class 3 stars.  A further 4743 were classified as field stars.
We distinguish here class 3 sources as defined in \citet{gut09}, which show weak IR emission consistent with a thin disk or weak line T-Tauri star, and 
the X-ray detected diskless class III sources, which do not show any IR  excess  emission and are classified as field stars (class 99) by the \citet{gut09} method.  

A comparison was made to the \citet{gut08} discovery paper.  All 92 YSOs reported in the discovery paper are matched to sources in the updated catalog, however 
the classifications have changed somewhat:  80/92 source classifications remain unchanged.  Two class 1 sources are now class 2, while a further two class 2 move to class 1.  
Another three class 2 sources are now stars (class 99). The remaining five objects (three class 1, two class 2)  are now unclassified.  The updated IRAC photometry 
 has improved on the initial classification.  We identify six new YSOs within the \citet{gut08} field of view, one class 1 and five class 2.  

A  comparison was made to the catalog of Serpens South members published by \citet{dunham}.  There were 268 of their sources located in our SS IR-field, of which 123/268 had 
a counterpart within 1'', with all 268 having a counterpart within 2''.    They employ a similar SED slope classification scheme as described Sec.~\ref{xrayysos}, 
 listing the $\alpha$ value in their catalog.  
Our $class_{color}$ classification method identified the 268 objects as 56 class 0/1, 138 class 2, 10 class 3, 50 field star, and 14 unclassified objects. 
Assigning a class based on their $\alpha$ gives:  47 class 0/I, 32 flat spectrum, 122 class II, 28 class III, and 39 field stars.  To simplify the comparison, we reclassify the flat spectrum sources as either class 0/I ($\alpha > 0$) or class II ($\alpha < 0$), giving 63 class 0/I and 138 class II . 
Comparing the two: 176 have the same classifications, while 37 show a difference of one class, 12 changed between disk-bearing and field star class, with the remaining not classed in one catalog.      Further, we matched the Dunham catalog to our X-ray source list, and found 19 matches using the our matching criteria.  This compares to 50 classified matches to our IR catalog, 33 if the field star class is excluded.     
In summary, of our 299 $class_{color}$ YSOs, 205 were found in the \citet{dunham} catalog and 87 were found in the \citet{gut08} catalog.  
Of the 299 YSOs, 64 were unique to our catalog, and are thus newly identified members of Serpens South.

\subsection{X-ray Luminous YSOs}\label{xrayysos}

The catalog of 152 X-ray sources was matched to the IR catalogue using a modulating radius of:
\begin{eqnarray*}
1'' + \pi (\theta/max(\theta))^2 
\end{eqnarray*}
, where $\theta$ is the off-axis angle of each X-ray source.  This provides a range in matching tolerances of 1''-4'' based on the distance of the source to the aim point.  
This accounts for the increased positional uncertainty and larger ellipse size with increasing off-axis distance in the ACIS-I image \citet{wol}.  
The positional accuracy of the $Chandra$ data is $\sim$0.5'' on axis and the positional accuracy of the updated IRAC catalog is considered to be better than 1'', therefore 
we set a minimum tolerance of 1'' in the centre of the ACIS field to ensure that mismatches are minimised in the dense core of Serpens South.    
Of the 152 sources, 66 were matched to an IR counterpart with one or more detections in the IR-bands. 

For the 66 IR-matched  X-ray detected YSOs a different approach was required from the \citet{gut2018} method to securely classify them,
  as the $class_{color}$ method can be insensitive to weak disk/class III sources.   Of these 66 sources, 35 were identified as YSOs with the $class_{color}$ technique from the IR photometry.   
The second method of classifying young stars from their IR photometry was to construct the Spectral Energy Distribution (SED), and calculate the slope, 
$\alpha = d log(\lambda F_{\lambda})/dlog(\lambda)$, over the mid-IR wavelengths.   

Protostellar objects (Class 0 and I) have a rising slope, $\alpha > 0.3$; flat spectrum objects have essentially flat slopes with $-0.3 < \alpha < 0.3$.  
Class II sources are characterized by decreasing slopes between$-1.6 < \alpha < -0.3$, while Class III sources lack any 
optically thick emission from a disk and possess decreasing slopes $\alpha < -1.6$, consistent with a stellar photosphere \citep{greene}.   
The dereddened slope of the SED over the four IRAC bands was used to further refine the evolutionary classification of each YSO, for those sources 
where the value of $A_K$ could be calculated   following \citet{gut05}.  The final classification was verified by visual assessment of the SEDs.   
 Of those 66, 21 are class I sources, 6 are flat spectrum, 16 are class II ojects, and 18 are class III diskless members of the cluster.   
 A further five could not be classified by this method; three were detected in only  one or two infrared bands and two were detected in 
three IRAC bands with ambigous SEDs.  

There are three major differences between our $class_{color}$ and $class_{SED}$ YSO classification schemes.   We include a separate flat spectrum class in order to 
match our  SED based classification scheme with the \citet{wolk2017} YSOVar paper on Serpens South and our previous Serpens Main work \citep{winston}. 
We combine the class 0 and class 1s into class I due to the lack of far-IR or submm photometry to securely differentiate between them.  
Finally, the X-ray detected class III sources were not identified as YSOs by the \citet{gut2018} method, where they are either '99: stars' or '-100: unclassified'.  
Their class 3 sources are considered to be weak class II objects according to the SED classification.  

The two classifications schemes provide very similar results:  the five unknown objects cannot be reliably classified by either scheme, leaving 61 
sources.  Of these, 42 (69\%) have the same classification (counting class III and class 99 as equivalent here ), 6 (10\%) are in the flat spectrum class, 8 (13\%) were not classified as YSOs by the $class_{color}$ method, and 5 (8\%) differed between protostellar and disk-bearing between the two methods.   
One X-ray source identified as a class I in the SED method and visually inspected in the mosaic, is classed as '9: shock' by the \citet{gut09} method. 

Fig.~\ref{fig2} shows two cmds: $H$v.$H-K$ and [3.6]v.[3.6-8.0], and the [3.6-4.5]v.[5.8-8.0] ccd.  
All sources in the field with photometric uncertainities $<$0.2 are shown as gray points.  The X-ray detected YSOs are overplotted as filled symbols:  
class I as red circles, flat spectrum as magenta hexagons, class II as green squares, and class III as blue stars.  
The [3.6]v.[3.6-8.0]  cmd shows that all the YSOs have a similar range in [3.6] magnitude with increasing color with earlier class as expected.  
The traditional IRAC ccd shows photospheric emission from the class IIIs, with one source perhaps showing weak outer disk emission. 

Fig.~\ref{fig3} shows histograms of four properties of the X-ray sources: the [3.6]~mag flux, the X-ray flux ($F_X$), the plasma temperature ($kT$), and the hydrogen column density ($N_H$).   
 The  X-ray detected YSOs cover a similar range as the full IR sample, indicating that there is no trend towards detecting only the brighter IR sources in X-ray emission.   
There are three groupings in the X-ray flux, with centers at $log(F_X)$~ -18, -14, -7;   sources in both the fainter and brighter groupings are spatially coincident with the 
center of the Serpens South cluster and those with IR counterparts are therefore likely to be bona fide low mass and flaring YSOs, though a number of brighter sources at  $log(F_X) \sim -6$ are possibly 
foreground dMe stars.   There is no trend in $N_H$ or $kT$ between sources with and without an IR counterpart.  This implies that we are missing a substantial number of 
cluster members in the IR; further higher resolution IR and X-ray observations are needed to investigate this.  Therefore, a high percentage of the 86 unmatched X-ray sources  may be  bone fide cluster members. 

The spatial distribution of the X-ray detected and IR-detected YSOs are presented in Fig.~\ref{fig4} on the background of the {\it Herschel} hydrogen column density map of the region.  
The image covers the approximate size of the SS IR field, with the white rectangle indicating the smaller {\it Chandra} field.  A discussion of the spatial distribution of the YSOs is 
presented in Section~\ref{spatial}.

\subsubsection{Contamination and Completeness}\label{contam}

Contamination of the X-ray source list from background extragalactic objects such as active galactic nuclei (AGN) and foreground and background 
galactic field stars (primarily dMe stars and dusty AGB stars) is expected. 
Such contamination is limited by requiring a photometric counterpart in the infrared and by comparison of modeled hydrogen column density, $N_H$, 
plasma temperature, $kT$,  and near-IR extinction, $A_K$, values: Foreground (and background stars) exhibit different colors to YSOs, have lower 
 (higher) extinctions, and are evenly distributed across the field.   AGN are fainter in the IR with higher extinctions and plasma temperatures, and a uniform 
distribution across the field.   

In a given {\it Chandra} field, with $\sim$100ks exposure, we would expect to detect $\sim$100 AGN contaminants. 
Statistical estimates of the populations of contaminating sources in {\it Chandra} observations of star-forming regions are discussed by \citet{getman2011} 
and references therein.  They report roughly 10-15\% of sources in both the ONC and Cep~B observations are contaminants, with 20-30 AGN and 15-20 
foreground stars in the shallow (30~ks) Cep~B survey, and 150-200 AGN and 10-15 foreground stars in the COUP $\sim$880~ks survey.  The stellar
background contaminants were negligble in each field.   A similar percentage in Serpens South would suggest that $\sim$10-20 sources of the 152 detected 
are foreground stellar contaminants.  Given the extremely high background extinction towards Serpens South, we expect that the contaminants would only be foreground 
stars and AGN, evenly distributed across the field of view.  For this reason it is highly unlikely that any of the YSOs are actually background dusty AGB stars.    
By requiring an IR counterpart for selection we reduce the likelihood of including background AGN 
contaminants in our sample of cluster members and can better distinguish older foreground stars based on color and extinction.  Many of the remaining X-ray 
detections, that are unmatched in the IR, are likely to be background AGN.   

In Figure~\ref{fig2},  the seven class III sources falling between the ZAMS and 1~Myr Siess isochrone \citep{sie} on the $H$v.$H-K$ cmd could be 
foreground contaminants, an older surface population associated with another cluster, or {\bf older} (more massive) young cluster YSOs.   
Four of the class III YSOs lie significantly below the reddening vector for an M6 star at 1~Myr, indictating that they are possibly very low mass stars or brown dwarfs, 
or  possibly contaminants.  The Chandra identifications of these sources are  ID\#31, 39, 42, 54. 
Given the high extinction at their positions close to the central filament ($N_H \sim 1.7-5.8\times10^{22}cm^{-2}$), they are 
unlikely to be behind the cluster, and  could therefore be foreground dMe stars or white dwarfs.   However, their photometry are not consistent 
with dMe dwarfs $J-H~v.~H-K$ colors \citep{lepine} or absolute magnitudes assuming they lie in front of the cluster \citep{riaz}.   
Similarly, they are not wholly consistent with the  $J-H~v.~H-K$ colors of white dwarfs,  with \citet{steele} showing their colors would require a stellar or brown 
dwarf companion and that they are unusually bright in the near-IR .  The four class IIIs $J-H~v.~H-K$ colors are consistent with young low mass dwarfs or brown 
dwarfs.   \citet{filippazzo} report absolute magnitudes for young  8-100Myr low mass dwarfs similar to the apparent magnitudes at 260-430pc to the cluster. 
Further, the on-sky spearation of the three faintest sources is 0.35-0.5pc if they are members of the cluster.  If they were foreground objects they would have 
an even smaller separation, which we suggest is unlikely.  Therefore, we retain these four objects as low mass class III members of Serpens South.

\section{\bf Discussion}\label{disc}

\subsection{X-ray Protostars \& Pre-Main Sequence}

We detect 21 class I and 6 flat spectrum protostars in X-rays, along with 16 class II pre-main sequence stars.    
This gives an X-ray protostellar fraction of 27/43 or 63$\pm$12\%. 
This fraction drops to 27/61 or 44$\pm$9\% when the class III sources are included.   
The former fraction is marginally higher than those found in Serpens Main (51$\pm$11\%) and 
NGC 1333 (49$\pm$8\%) \citep{winston, win09}, which is in agreement with Serpens South being considered the youngest of the three regions, 
though the differences lie within the uncertainties.

The X-ray detection fraction by evolutionary class is an approximation due to the requirement for an uncertainty less than 0.2~mag in the IR selection, 
which is not required for the X-ray detected YSOs, and the increased psf with off-axis angle.   
With this in mind, the detection fractions for  for X-ray detected YSOs to all X-ray and IR detected YSOs 
are 27/57 (47$\pm$9\%) for class 0/I (including flat spectrum), and 14/69 (20$\pm$5\%) for class II, 
for sources with the ACIS-I FOV, as listed in Table~\ref{sid}.      
These percentages are not inconsistent with half of each class being detected as has previously been seen in Serpens Main \citep{winston}, 
for the protostellar sources.  This would imply that the X-ray generation mechanism is not inhibited at the earlier stages of a cluster's development.  
However, the fraction of class II sources detected is much lower than half; the class II population is more widely distributed over the 
{\it Chandra} field of view, meaning that a higher fraction of the class IIs are located at high off-axis angles, where the sensitivity and positional 
accuracy of the X-ray data decreases.  
To account for these effects, we examined the fraction in the central Serpens South core region defined as being a 6~arcmin long segment of the 
filament centred on the {\it Chandra} pointing; here we find 6/17 (35$\pm$14\%) for the class II sources.  

Two of the sources, \#28 and \#98, were matched to the locations of detections VLA7 and VLA17 in \citet{kern}, respectively.   Source \#28 is a class II variable object,  
while source \#98 is a class I object and was not found to be variable.      A further two sources, \#14 and \#72, were matched to within 10'' of two of the 1.2mm MAMBO 
sources detected by \citet{maury}: SerS-MM12 and SerS-MM15, respectively.   However, neither X-ray source has a counterpart in the IR.

\subsection{X-ray Class III Diskless YSOs}

We have detected eighteen class III X-ray sources in Serpens South.   As can be seen in Fig.~\ref{fig4}, six of these sources are centred directly on the cluster 
core within 2.5' of the aimpoint.   It is unlikely that they are all foreground or background contaminants: 
they do not meet the IR criteria for extragalactic contaminants, are brighter in IR than would be expected of background stellar contaminants, and we estimate only 
a few foreground stars at the distance of Serpens South and the length of the exposure.   Further, their spatial distribution does not appear to be random across the field.  
It is thus likely that the sources are bona fide diskless young members of Serpens South.   Their location in the center of the cluster, a region still dominated by protostars, 
strongly suggests  that they are not an older population, but very young stars that have processed their disks at an accelerated rate.  
It is likely that external environmental factors have played a role:   
Such effects as disk stripping by more massive neighbors or loose binaries, tidal disruption via close encounters, and ejection from a multiple system may 
have lead to these diskless young stars. 
A further four class III objects are located adjacent to the Serpens South core, while four more are located along the filament visible as shown in Figure~\ref{fig4}.  
The remaining four are located to the west of the core, are brighter in IR magnitude though not X-ray flux, and may be associated with another region to the west, c.f. Sec.~\ref{spatial}.  

The X-ray diskless fraction in Serpens South is found to be 18/34 or 53$\pm$12\% for the class II and III sources, and 18/61 or 30$\pm$7\% 
when the protostellar sources are included.   Allowing for the wide variation, this represents a large fraction of diskless members for the $<$1~Myr 
cluster, and is similar to the $48\pm$11\% fraction found in Serpens Main.   
These values represent an upper limit to the diskless fraction in the field due to the incompleteness of the combined X-ray and IR sample.  

Another approach is to estimate the disk fraction of the cluster and thus estimate the cluster's age, as originally presented by \citet{her2}.  
For Serpens South, the X-ray disk fraction ranges from a minimum of 43/61 or 70$\pm$11\% for the X-ray detected YSOs, 
to 29/33 or 88$\pm$16\% for YSOs located directed in the Serpens South core.   This represents a lower range of values than would be expected for such a young cluster.   

One implication of the presence of class IIIs in this cluster may be that many young stars are essentially 'born' with little to no circumstellar disk and/or 
envelope material, and that the population of class III objects which we have so far examined is not likely to be an older, more evolved population as is 
generally assumed, but is instead young and coeval with the class I and II population of the cluster. This conclusion is borne out by the similarities in the 
properties between the median fluxes and plasma temperaturesof the two groups. 
Their presence in the youngest known nearby cluster questions the assumption that there is a direct relation between disk fraction and cluster age.  
The cluster environment: stellar density, stellar distribution, cloud dispersal, may all play larger roles than previously assumed \citep{pfalzner}.  
Three of  the class III objects are also likely to be very low mass stars/brown dwarfs, possibly implying they were ejected from a multiple system at an early 
stage of their formation.     
If disk lifetimes have been affected by external factors, this would lead to estimates that tend towards shorter timescales.   

In our previous spectroscopic work on Serpens Main and NGC 1333 it was found that the class III population was coeval with (and in some cases apparently younger than) 
the class II population in these $\sim$1 Myr clusters \citep{win09, win10}. The similar result in Serpens South indicates that the identified class III population is, 
perhaps, not older or more evolved but may be a product of the active and dynamic cluster environment. A similar spectroscopic survey of Serpens South is currently 
underway to determine the isochronal ages of the members of this young cluster. Further large scale surveys of these regions will be necessary to finally identify any 
older populations of evolved class III young stars if we wish to truly examine the temporal evolution of these young star forming regions.

\subsection{X-ray Spectral Properties}

The $N_H$v.$kT$ relation, the upper left plot of Fig.~\ref{fig5}, shows a roughly linear trend of increasing column densities with increasing plasma temperatures across 
all classes.  An $OLS$ fit to the data gives a relation of $N_H\sim0.8555\pm0.174~kT$ .   The more evolved classes show lower average $N_H$ values, as expected for less embedded objects.   
The relation of $F_X$ to $kT$, the middle left plot of Fig.~\ref{fig5}, was fitted using an OLS routine and shows a weakly decreasing trend of 
$log(F_X) \sim -0.147\pm0.096~kT  - 12.768\pm0.202 $ . This is due partly to the wide range in $F_X$ at lower values of plasma temperature.  
The middle right plot of Fig.~\ref{fig5} shows the relation of  $F_X$ to $N_H$, with a fit of $log(F_X) \sim 0.0479\pm0.050~N_H - 13.277\pm0.156$ . 
There is some evidence of a trend towards lower fluxes and higher column densities as the YSOs move to earlier evolutionary classes.  
To determine if there was a trend in X-ray flux with distance from the center of the cluster, we compared $F_X$ to the off-axis angle, $\Theta$, and 
found a weak trend of $log(F_X) \sim 0.0495\pm0.04 ~ \Theta - 13.262\pm0.263$ , where the X-ray flux has a slight tendency to increase the further 
off-axis the source is located.  This is likely due to the location of the fainter protostellar sources near to the center of the aim-point.

There is a weak trend in $F_X$ with dereddened H-band magnitude observed, c.f. lower left of Fig.~\ref{fig5}, with 
$log(F_X) \sim -0.126\pm0.093~[H]_{dered} - 11.875\pm0.982$ .  
The dereddened H-band serves as a proxy for stellar mass as it is the band least contaminated by reddening and emission 
from the circumstellar material (though these are not completely removed and will contribute to scatter in the plot).  
Given that the X-ray flux is a function of bolometric luminosity, which is itself a function of mass, such a trend is expected though 
with a large scatter due to the large variation in rotational periods \citep{gall}.  A similar weak trend was reported in Serpens Main 
and LkHa~101 \citep{winston, wol2}.

Fig.~\ref{fig5} (upper right) shows the $N_H$v.$A_K$ relation, which compares the density of the gas to extinction from dust in the line of sight.  
Historical measurements of $N_H/A_V$ range from approximately $2.2\times 10^{21} cm^{-2}$ \citep{ryt}  
derived from O-star absorption, to roughly $1.6\times 10^{21} cm^{-2}$ \citep{vuo} derived from a well 
behaved sample of PMS stars in the $\rho$ Ophichus cluster. 
The gas to dust ratio in Serpens South was fitted as $N_H$$\sim$$0.6845\pm0.094 \times 10^{22}~A_K$.  
We have included class II and flat spectrum sources in the fitting due to 
the paucity of available class III diskless objects, which best trace the intra cluster medium, especially at higher values of both $N_H$ and $A_K$ .        
A lowering of the dust to gas ratio can be due to depletion of hydrogen gas or to grain growth or annealing into crystalline silicates \citep{winston, jura}.   
This value is consistent with those of the lower mass clusters Serpens and NGC 1333, where the authors reported on a 
decreased $N_H$ to $A_K$ ratio, of $\sim0.6 \times 10^{22}$ \citep{winston,win10}, than that 
quoted for the diffuse interstellar medium in \citet{vuo} of  $1.6 \times 10^{22}$ .   
However, the high-mass RCW~38 and RCW~108 fit very well to the ratio of $1.6 \times 10^{22}$, consistent with the \citet{vuo} value for the local 
ISM and nearby molecular clouds \citep{win11, wol11}.

\subsection{X-ray Luminosity and Cluster Distance}

The X-ray luminosity function (XLF) for young clusters has been examined by \citet{fei} and demonstrated empirically to follow a 'universal' log normal distribution, 
where $<log(L_X)> = 29.3$  and $\sigma = 1.0$.     
This univesal XLF is dependent only on the assumed distance to the cluster, making it a useful proxy for estimating the distance to these young clusters.
 While this method is nominally independent of the optical or IR photometry of the cluster, it is dependant on stellar age and spectral type.  The most massive 
stars are in general the most X-ray luminous and as such set the high energy tail of the XLF distribution.     

 The authors have previously presented the XLF for Serpens Main \citep{win11}, showing that it is better fit by a distance closer to $\sim$360pc, 
than to the then accepted distance of 260pc \citep{str}.  This was closer to the current best distance measurement of 436$\pm$9pc to that region \citep{ortiz}.   
The distance to the Serpens Cloud and the Aquila Rift complex has been the subject of much debate in the literature.  
Serpens South is assumed to lie at the distance of Serpens Main, however this has not been verified.    

An early distance determination to the Serpens Main region was that of \citet{strom} who measured the distance to HD 170734 as 440 pc, 
assuming A0 spectral type, V = 9.2, and E(B-V)$\sim$0.3. \citet{zhang} reported distance of 700 pc was based on HD 170634, HD 170739, 
and HD 170784 with B spectral types, and R = 3.1. \citet{delara} used the same stars, adding BD-24607 and Chavarria 7, reclassifying them 
to be on the main sequence, 
with R$_{BV}$ = 3.3$\pm$0.3 and obtained a distance of 310 pc. \citet{str} detail Vilnius photometry and photometric 
classification of 473 stars toward the Serpens Cauda cloud complex, to a depth of V$\sim$13. The $A_V$ and distance to each star 
were calculated and used to determine the near edge of the Serpens Main cloud and to estimate the depth of the cloud. The near 
edge of the cloud was found to lie at 225$\pm$55~pc, with an estimated depth of 80~pc, average distance of 260~pc, and far edge at 360~pc.
A recent distance estimate to Serpens is that of \citet{dzib2010} who used the VLBA to measure the parallaxes of both components of 
the YSO EC~95, which is located at the centre of the Serpens Main cluster.  They estimate a distance of 429$\pm$2~pc.   
\citet{shup} use spectral typing of MS stars to estimate a distance of 455-535~pc to W40, which lies adjacent to Serpens South.   
This is consistent with the result of \citet{ortiz} of 436~pc from VLBA measurements.   

We have here undertaken a similar XLF analysis for Serpens South, however in this case, instead of attempting to assign an exact distance based 
on the empirical function, we compare the XLFs of three clusters:  Serpens Main, NGC~1333, and W~40.   This provides a comparitive distance 
measure and the likelihood of an association between Serpens South and its two nearest neighbors, Serpens Main and W~40.  
We include a comparison to NGC~1333, as its distance of 240~pc is considered to be reliable, and exhibits an XLF strongly consistent with this distance.  
The X-ray datasets for the Serpens Main and NGC~1333 clusters were obtained from the Chandra ANCHORS archive \footnote{http://cxc.harvard.edu/ANCHORS/}.  
The W~40 luminosity values were taken from the \citet{kuhn} paper, and then adjusted from 600pc to a distance of 430pc.

In Figs.~\ref{fig6}~\&~\ref{fig7} the histogram and cumulative distributions of the X-ray luminosities 
are compared for the currently accepted distances to Serpens Main (430pc), NGC~1333 (240pc), and at a distance of 430~pc for W~40.  W~40 has recently 
been found by VLBA to lie at the same distance as Serpens Main, a far closer distance than the previously accepted range of 600-900pc \citep{ortiz}.  
If all three clusters are associated this would make the Aquila Rift complex a massive star forming complex lying at essentially the same distance as Orion.  

Following \citet{2016ApJ...820L..28P} we compared the median X-ray luminosity of Serpens South over a range in distances to the medians of Serpens Main 
and NGC~1333 at their accepted distances to determine the distance at which they are similar.  The median value for Serpens Main was $log(L_X)\sim29.77$, 
and for NGC~1333 was $log(L_X)\sim29.73$, for $log(L_X) > 29.3$.  The median flux for Serpens South at 430~pc was $log(L_X)\sim30.15$, and $log(L_X)\sim29.77$ 
at 260~pc.    We also applied the Kolmogrov-Smirnov (K-S) test to determine the probability that two samples are drawn from the same distribution.  
The highest probability that the NGC~1333 and Serpens South samples were taken from the same distribution ([K-S stat: 0.1,  prob.: 0.996]) occurred at a 
distance of 260pc for Serpens South.  The results were  [K-S stat: 0.48,  prob.: 0.0005] at a distance of 430pc to Serpens South.    
Similarly, the highest probability that the Serpens Main and Serpens South samples were taken from the same distribution ([K-S stat: 0.147,  prob.: 0.880]) 
also occurred at a distance of 260pc.  The results were [ K-S stat: 0.437,  prob.: 0.0041] for Serpens South at a distance of 430pc.   

We also compared W~40 with NGC~1333, and found that the median values of W~40 were $log(L_X)\sim29.67$ and $log(L_X)\sim29.79$ at 430pc and 600pc, respectively. 
The median value at 600~pc is closer to the NGC~1333 median, in agreement with the \citet{kuhn} distance estimate to the cluster.   
However, the K-S test results are only marginally better fit by the further distance, with  [K-S stat: 0.139,  prob.: 0.659] and [K-S stat: 0.134,  prob.: 0.650] 
for 430pc and 600pc, respectively.   For this reason we use the \citet{ortiz} distance in Figs.~\ref{fig6} \& \ref{fig7}.       

Surprisingly, then, the best fit distance to Serpens South when compared to the XLFs of the other regions is 260pc, one of the nearest distance estimates to the 
Aquila Rift cloud complex  \citep{str}.    This implies that the cluster is closer than Serpens Main and W~40, lying in front of these two regions, and that it is not 
likely to be physically associated with either region.   It is possible that a number of the brighter sources (on which the fitting rests) are from a foreground 
population near the 'surface' of the Aquila cloud and that the main young cluster is embedded further into the cloud at a more similar distance to the other two regions.  

Further planned observations with $XMM-Newton$ of extended fields surrounding Serpens South will help clarify this issue by identifying any distributed populations or a
trend in the X-ray properties that may indicate contamination from different cluster populations.  
Future $Gaia$ releases may also provide distances to some of the brighter and less embedded X-ray sources.  Gaia will, however, suffer several problems evaluating such very 
young clusters: its cut-off at $V\sim20$, making detections towards dense cores with high $A_V$ difficult.  Gaia may provide reliable distances to the less embedded and more evolved objects - the Class IIIs detected by {\it Chandra} - the current release did not contain any parallax measurements for known cluster members of Serpens South.

\subsection{Spatial Distribution}\label{spatial}

The Serpens South core cluster was first identified in IRAC imaging by \citet{gut08} as a knot of bright stars in the center of a dusty filament.  
However,  an extended population of young stars exists in the region, merging into W40 in the east.  
To better understand the YSO spatial distribution, we have examined the IR-excess population of YSOs surrounding the {\it Chandra} field of view and have overplotted 
their spatial distribution on the {\it Herschel Gould's Belt} survey column density map of the region, Fig.~\ref{fig4} \citep{kony}.    
The hydrogen column density map shows a dense filament extending from northwest to south where it splits into two lanes.  
There is another more fragmentary filament to the northeast, and a highly dense spherical region to the west.  These regions show peak column densities of 
$\sim2\times10^{22} cm^{-2}$.  They are also the locations of the majority of the protostellar population.  The more evolved disk-bearing and diskless stars are 
more widely distributed, though both groups still trace the filament and high density regions.  

The large number of class II objects to the southeast and southwest of the main filament may indicate the presence of an older population, perhaps associated 
with W40 in the east, and the small dense cluster to the west.  The high percentage of class IIs and class IIIs relative to the protostars would imply that these groups 
are more evolved, and as such older, that the central Serpens South cluster.

As an initial approach to understanding the spatial distribution of the IR-detected YSOs in the region, a surface density map is shown in Figure~\ref{fig8}.   
The local surface density at each point on a uniform grid was calculated following the method 
outlined in \citet{case}:
\begin{eqnarray*}
\Sigma(i,j) = \frac{N-1}{\pi~r_{N}^{2}(i,j)}
\end{eqnarray*}
\noindent where $r_N$ is the projected distance to the $Nth$ nearest cluster member.  In this case, the stellar surface density was calculated with a grid size of 
$100\times100$ using $N = 18$, the distance to the 18th nearest neighbor, to smooth out smaller scale structure.  
Further, \citet{case} found that the uncertainty goes as $\Sigma/(N-2)^{1/2}$, so by taking $N=18$, the uncertainty is 25\%. 
The underlying filamentary structure is clearly visible, with the Serpens South core and SE cluster forming the two highest density areas.   
The cluster to the west also forms a clear peak in stellar density, while a tentative cluster to the NE is faintly visible in the log-scaled image.

To perform a quantative search for substructure in the Serpens South IR-field a minimum spanning tree (MST) graph was calculated for the class 0/1 
and class 2 sources, as shown in Fig.~\ref{fig9}.   This graph connects the YSOs in such as way as to connect each point without loops while minimising the total 
length of the branches, and was done using a {\it python} routine written by \citet{vand}.   Subclusters were identified as those points connected by branches with 
lengths less than the characteristic branch length \citep{gut08}.  This is defined from Fig.~\ref{fig10} as the intersection of linear fits to the cumulative 
distribution of branch lengths; for the Serpens South IR-field is 0.024$^o$ or 87$''$.  This branch length corresponds to on sky separations of 0.11pc and 0.18pc, at 
cluster distances of 260pc and 430pc, respectively.   Four subclusters are identified using this method.   
Three are associated with the central filament; two smaller clusters at each end and the main Serpens South core at its center.    
The fourth subcluster is associated with the dense region to the west and is composed 
primarily of pre-main sequence stars, indicating that it is either an older cluster or lies behind an older foreground population.   
The tentative subcluster is associated with the more fragmented, possibly ring-shaped,  filament to the northeast. 
This cluster contains only about twenty members, of which six are protostars, 
a similarly high fraction to the central cluster, indicating it may be of a similar age though with far fewer members.  
 It is not identified as a subcluster in the MST approach as we require a minimum of 10 objects to be assigned to a subcluster and only six are in close enough proximity 
due to the apparently ring-shaped distribution.       

Subclustering in a star-forming region can be further quantified using the Q-parameter as defined by \citet{car}:  it extends the usage of the MST to quantify and 
distinguish between smooth large-scale radial density gradients and multiscale or fractal subclustering. 
This method provides a statistical method of characterizing structure in stellar clusters.  The $Q$-parameter is defined as $\bar{m} / \bar{s}$, 
where $\bar{m}$ is the normalized mean MST branch length, $\bar{m}/(N_{total}A)^{1/2}/(N_{total}-1)$, and $\bar{s}$ is the normalized mean separation,  
$\bar{s} / R_{cluster}$. A $Q$-parameter equal to $1$ indicates a smooth radial distribution of sources, while lower values indicate subclustering.  
For the Serpens South IR-field, $\bar{m} =$ 0.0269$^o$ and $\bar{s} = $ 0.0365$^o$, yielding a value of $Q$ equal to 0.737.  This value is significantly less 
than 1 and indicates that fractal subclustering defines the structure and spatial distribution of the YSOs in Serpens South.  
This is in agreement with the visual association of the YSOs with the dusty filamentary structures as shown in Fig.~\ref{fig4}.

Overall,  it appears that the Serpens South region is dominated by a dense filament wherein the Serpens South core forms the central densest region of star formation 
with ongoing star formation along the entire length.   A small pocket of star formation to the NE in a less dense filament may be associated with 
Serpens South or the nearby W40 region.  The cluster to the west, with a lower protostellar fraction, may be slightly more evolved and its association with 
Serpens South, though likely, cannot be confirmed.  The X-ray detected class IIIs to the west of the Serpens South core are coincident with {\bf class 2} YSOs associated 
with this cluster, and therefore may belong to it and not the Serpens South core.

\subsection{Variables \& Flares}

The lightcurves of the 152 X-ray detections were searched for variability using the Gregory-Loredo based algorithm {\it CIAO} routine, {\it glvary}.  This routine 
returned the binned lightcurves and a probability index from 0-10, where zero is non-variable and $>$7 indicates a probability $>$0.9 that the source is variable.    
Of the 152 sources, 16 have $P>0.9$ to be variable, and 2 have $0.66<P<0.9$ to be variable.  
The lightcurves of the 18 variable sources are shown in Fig.\ref{fig11}, labeled with 
the source ID in  Table~\ref{IRphot1}  and their probability. 
Of the 18 variables detected: 4/21 class I are found to be variable, 1/6 flat spectrum are variable, 5/16 class II show variability, and 5/18 class III sources are variable.   
The remaining three did not have match in the IR catalogue.  Given the high extinction towards the cluster it is likely that they are cluster members
 undetected in the IR.   
A similar number of protostars and pre-main sequence cluster members are found to be varying, indicating little difference between the evolutionary classes.   

Eight sources exhibit flaring activity during the 97.5~ks observation.  There is no trend with class and flaring activity, and all the flares show a similar increase, 
of roughly 5 times the basal flux, at their peak emission.  
Seven sources exhibit a monotonic decrease and two a monotonic increase in activity during the observation.  

One unusual source, ID\#58,  is a class III source showing no excess emission and is found to be periodic at X-ray wavelengths.  It has a period of $\sim$0.463~days 
and shows a doubling in flux at peak levels.   
A number of explanations for this periodicity are possible:  stable spots on the stellar surface,  a binary companion, or the presence of a massive exoplanet.  
A low mass binary companion or a close-in hot jupiter-like exoplanet, that periodically 
rotates into view could block the X-ray emission of the primary.  \citet{flaccomio2005} report the rotational modulation 
of 23 X-ray sources in the COUP survey, finding 20-70\% amplitude changes on period or half-period timescales.  
Unfortunately, this source lies outside the \citet{wolk2017} YSOVar field of view and so no data is 
available on its variability at IR wavelengths.   However, approved observations of the extended Serpens South region 
with {\it XMM-Newton} will provide further coverage of this object and provide a better understanding of the periodic nature of this source.

\section{Summary}\label{summ}

We have undertaken a {\it Chandra} X-ray study of the Serpens South star forming cluster. Combining the X-ray data with near and mid-IR datasets from {2MASS and {\it Spitzer} we have identified  95 new YSOs and reassessed the membership of others.  The X-ray data set has allowed us to search for more evolved, class III members of this extremely young cluster to examine the extent of disk evolution and/or processing that has occurred in this young region.

The location of five of the class III sources in the center of the Serpens South cluster suggests that dynamical evolution can play a strong role in the processing of circumstellar disks, and therefore that cluster age estimates based on disk fractions are likely misleading.  

\begin{itemize}

\item We have identified 152 X-ray sources in the field, with 66 matched to an IR counterpart.  

\item Of the 66, we classify 21 class I, 6 flat spectrum, 16 class II, and 18 class III sources  with the class$_{SED}$ method. 
Five were detected in only one/two IR bands, and could not be classified.  Of these, 31 were new X-ray detected YSOs.  

\item The gas-to-dust ratio as measured by the $N_H$v.$A_K$ relation was found to be $N_H$$\sim$$0.68 \times A_K$, similar to Serpens Main, but lower than 
expected for the ISM, likely due to grain growth.    

\item The Serpens South X-ray luminosity function was compared to that of the Serpens Main and W40 clusters.  Our results indicate that it does not lie at the same 
distance as either cluster, with a best match to the nearer estimate of 260~pc for the front of the Aquila Rift complex.   

\item We report 299 YSOs with IR-excess emission, in a 0.6$^o$ square field centred on the {\it Chandra} Serpens South pointing; of these 90 were class 0/1 protostars, 
195 class 2 objects, and 14 weak disk-emission class 3  young stars  using the class$_{color}$ method.   Of these, 64 were newly identified YSOs in the IR.  

\item The protostars are all located in regions of high column density, as traced by {\it Herschel}.  The class II sources are somewhat more distributed over the field, 
however they still trace the underlying filamentary structure, which is consistent with the extremely young age of Serpens South.   

\item Five of the class III sources are coincident with the central core of the Serpens South cluster, suggesting that the evolution of their disks has occured not due to age, 
but to dynamical evolution between young stars in the cluster.   

\item Four subclusters were identified in the IR-field using the MST technique, with a $Q = 0.737$, indicating fractal subclustering. 
The surface density map shows that the three central subclusters are knots of star formation along the central filament.   
The cluster to the W may not be associated with Serpens South; it appears to be older, with a lower protostellar fraction, and may be 
the origin of 4-6 of the class III stars identified to the W of the Serpens South core.      
A fifth smaller subcluster to the NE may be simply be an overdensity or a could be a much smaller/dispersed ring-shaped cluster.    

\item We found 18 variable sources, in all evolutionary classes, 8 of which show flares.  One class III source showed periodic variability with a 0.463~day period.

\end{itemize}

We would like to thank the anonymous referee for their detailed review and helpful comments that have greatly improved the manuscript.   

This work is based on observations made with the {\it Chandra} Telescope (OBSID 11013). 
This work is based on observations made with the Spitzer Space Telescope (PID~30574), which is operated by the Jet Propulsion Laboratory, California Institute of Technology 
under NASA contract 1407. 
R. Gutermuth gratefully acknowledges funding support for this work from NASA ADAP grants NNX11AD14G, NNX13AF08G, NNX15AF05G, and NNX17AF24G.
This research has made use of data obtained from the Chandra Data Archive and the Chandra Source Catalog, and software provided by the Chandra X-ray Center (CXC) in the 
application packages CIAO, ChIPS, and Sherpa.
This publication makes use of data products from the Two Micron All Sky Survey, which is a joint project of the University of Massachusetts and the Infrared Processing and 
Analysis Center/California Institute of Technology, funded by the National Aeronautics and Space Administration and the National Science Foundation. 
Support for the IRAC instrument was provided by NASA through contract 960541 issued by JPL.

\clearpage


\begin{deluxetable}{cccc}
\tabletypesize{\scriptsize}
\tablecaption{Table of Identified Cluster YSOs.\tablenotemark{a}  \label{sid}}
\tablewidth{0pt}
\tablehead{\colhead{Class}  &  \colhead{Ser S IR-field}  & \colhead{$Chandra$ FOV}   & \colhead{Detected by $Chandra$}  }
\startdata
C I /  C 0 1   &  90  &  57  &  21 \\
FS   &   -  &  -  &   6 \\
C II/ C 2   &  195  &  66  &  16 \\
C 3 & 14  & 3  &  -  \\
C III   &  -  & -   &  18 \\
UNK   & - & -  &  5  \\
\hline
Total  &  299  &  126  &  66  \\
\enddata

\tablenotetext{a}{The numbers of YSOs identified by both methods in the two fields outlined in Figure~1.  The numbers detected by Chandra include YSOs not classified as such by our IR YSO identification.  The 'C 3' category are those objects identified as anaemic-disk class III via the $class_{color}$ classification method but which were not detected in X-rays.   }

\end{deluxetable}

\center
\begin{deluxetable}{ccccccccccccccccccc}

\tablecolumns{19}
\rotate
\tabletypesize{\scriptsize}
\setlength{\tabcolsep}{0.03in}
\tablewidth{0pt}
\tablecaption{{\it Chandra} X-ray Detections \label{IRphot1}}

\tablehead{ \colhead{Ch.} & \colhead{IR \tablenotemark{a}} & \colhead{RA} & \colhead{Dec} & \colhead{Src.\tablenotemark{b}} & \colhead{Off} & \colhead{Net} & \colhead{$\sigma$Net} & \colhead{N$_H$} & \colhead{N$_{H, l}$} & \colhead{N$_{H, u}$} & \colhead{kT} & \colhead{kT$_{l}$} & \colhead{kT$_{u}$} & \colhead{Abs. F$_{X}$} & \colhead{Abs. $\sigma$F$_{X}$}  & \colhead{F$_{X}$ \tablenotemark{c}} & \colhead{$\sigma$F$_{X}$}   & \colhead{Var.\tablenotemark{d}} \\

\colhead{ID} &\colhead{src?}  &\colhead{} & \colhead{} & \colhead{Sig.} & \colhead{Axis} & \colhead{} & \colhead{} & \colhead{$10^{22}$} & \colhead{$10^{22}$} & \colhead{$10^{22}$} & \colhead{} & \colhead{} & \colhead{} & \colhead{$ergs$} & \colhead{$ergs$} & \colhead{$ergs$} & \colhead{$ergs$} & \colhead{}    \\

\colhead{} &\colhead{} & \colhead{(J2000)} & \colhead{(J2000)} & \colhead{} & \colhead{deg.} & \colhead{Cnts} & \colhead{Cnts} & \colhead{$cm^{-2}$} & \colhead{$cm^{-2}$} & \colhead{$cm^{-2}$} & \colhead{keV} & \colhead{keV} & \colhead{keV} & \colhead{$cm^{-2}$$s^{-1}$} & \colhead{$cm^{-2}$$s^{-1}$} & \colhead{$cm^{-2}$$s^{-1}$} & \colhead{$cm^{-2}$$s^{-1}$} & \colhead{}  

}

\startdata

1 & IR & 18h29m15.977s & -2d08m36.422s & 8.10 & 13.23 & 196.33 & 24.19 & 1.73 & -0.25 & 0.25 & 1.74 & -0.26 & 0.26 & 3.25e-14 & 3.17e-14 & 1.17e-13 & 3.23e-14 & ... \\
2 & IR & 18h30m09.2883s & -2d07m23.1351s & 9.08 & 5.98 & 91.14 & 10.02 & 2.14 & ... & ... & 1.83 & ... & ... & 1.62e-14 & 1.59e-14 & 6.19e-14 & 1.56e-14 & V \\
3 & ... & 18h30m08.6374s & -2d06m09.6011s & 8.04 & 4.77 & 80.49 & 9.99 & 4.57 & ... & ... & 2.32 & ... & ... & 1.82e-14 & 1.61e-14 & 8.04e-14 & 1.69e-14 & ... \\
4 & ... & 18h29m44.1198s & -2d05m20.3177s & 2.31 & 5.58 & 14.12 & 6.07 & ... & ... & ... & ... & ... & ... & 3.05e-33 & -- & 7.86e-21 & -- & ... \\
5 & IR & 18h30m03.136s & -2d05m14.8525s & 12.31 & 3.52 & 154.63 & 12.53 & 1.96 & ... & ... & 2.25 & ... & ... & 2.53e-14 & 2.48e-14 & 7.68e-14 & 2.49e-14 & V \\
6 & IR & 18h29m03.4089s & -2d04m51.5s & 3.80 & 14.78 & 37.23 & 9.77 & 0.74 & -0.35 & 0.35 & 0.99 & -0.20 & 0.20 & 7.49e-15 & 6.86e-15 & 3.06e-14 & 7.16e-15 & ... \\
7 & ... & 18h30m21.1888s & -2d04m30.0481s & 5.19 & 5.69 & 40.90 & 7.86 & 15.23 & ... & ... & 31.10 & ... & ... & 2.24e-14 & 2.23e-14 & 7.42e-14 & 2.24e-14 & ... \\
8 & IR & 18h30m04.4512s & -2d04m15.8764s & 9.11 & 2.63 & 86.68 & 9.49 & 2.96 & -0.45 & 0.45 & 2.72 & -0.45 & 0.45 & 1.68e-14 & 1.65e-14 & 5.29e-14 & 1.67e-14 & V \\
\enddata

\tablenotetext{a}{IR src: X-ray source has IR counterpart. }
\tablenotetext{b}{Src. Sig.: Source Significance. }
\tablenotetext{c}{$F_{x}$: corrected unabsorbed flux }
\tablenotetext{d}{Var: indicates source is X-ray variable (V) or possibly/weakly variable (wV). }

\end{deluxetable}

\center
\begin{deluxetable}{cccccccccccccccccccc}
\tablecolumns{20}
\rotate
\tabletypesize{\scriptsize}
\setlength{\tabcolsep}{0.03in}
\tablewidth{0pt}
\tablecaption{Infrared Photometry of X-ray Detections in Serpens South \label{IRphot}}

\tablehead{ \colhead{Ch.} &\colhead{ J} & \colhead{unc.  J} & \colhead{H} & \colhead{unc. H} & \colhead{K} & \colhead{unc. K} & \colhead{3.6} & \colhead{unc. 3.6} & \colhead{4.5} & \colhead{unc. 4.5} & \colhead{5.8} & \colhead{unc. 5.8} & \colhead{8.0} & \colhead{unc. 8.0} & \colhead{24} & \colhead{unc. 24} & \colhead{A$_K$} & \colhead{Class$_{SED}$} & \colhead{Class$_{Color}$} \\
\colhead{ID} &\colhead{mag} & \colhead{mag} & \colhead{mag} & \colhead{mag} & \colhead{mag} & \colhead{mag} & \colhead{mag} & \colhead{mag} & \colhead{mag} & \colhead{mag} & \colhead{mag} & \colhead{mag} & \colhead{mag} & \colhead{mag} & \colhead{mag} & \colhead{mag} & \colhead{mag} & \colhead{} & \colhead{} 
}

\startdata

1 & 14.24 & 0.07 & 12.48 & 0.05 & 11.46 & 0.05 & 9.84 & 0.00 & 9.32 & 0.00 & 8.94 & 0.00 & 8.30 & 0.01 & ... & ... & 1.01914 & II & 2 \\
2 & ... & ... & 13.71 & 0.04 & 11.77 & 0.03 & 9.96 & 0.00 & 9.26 & 0.00 & 8.67 & 0.00 & 7.92 & 0.00 & 5.33 & 0.01 & 2.10828 & II & 2 \\
5 & ... & ... & 14.72 & 0.07 & 12.66 & 0.03 & 11.06 & 0.00 & 10.53 & 0.00 & 10.20 & 0.01 & 9.73 & 0.02 & 6.14 & 0.03 & 2.93035 & II & 2 \\
6 & 14.26 & 0.03 & 12.96 & 0.03 & 12.39 & 0.03 & 11.75 & 0.00 & 11.46 & 0.00 & 10.97 & 0.01 & 10.21 & 0.02 & 7.46 & 0.10 & 0.710205 & II & 2 \\
8 & ... & ... & ... & ... & 13.49 & 0.06 & 10.94 & 0.01 & 9.83 & 0.00 & 8.83 & 0.00 & 8.04 & 0.00 & 5.73 & 0.03 & ... & FS & 2 \\
9 & 15.40 & 0.07 & 12.98 & 0.03 & 11.93 & 0.03 & 11.21 & 0.01 & 11.08 & 0.01 & 10.92 & 0.01 & 10.91 & 0.04 & ... & ... & 1.90737 & III & 99 \\
10 & ... & ... & ... & ... & ... & ... & 12.70 & 0.01 & 10.93 & 0.00 & 9.99 & 0.01 & 9.42 & 0.02 & 5.92 & 0.02 & ... & FS & 1 \\
13 & ... & ... & ... & ... & ... & ... & 10.34 & 0.03 & 9.27 & 0.06 & 8.43 & 0.04 & 7.54 & 0.03 & ... & ... & ... & I & 1 \\
\enddata
\end{deluxetable}

\center
\begin{deluxetable}{ccccccccccccccccccccc}

\tablecolumns{21}
\rotate
\tabletypesize{\scriptsize}
\setlength{\tabcolsep}{0.03in}
\tablewidth{0pt}
\tablecaption{ IR Properties of identified YSOs in Serpens South IR-field \label{IRphotYSOs}}

\tablehead{ \colhead{IRAC} &\colhead{ RA} & \colhead{Dec} &\colhead{ J} & \colhead{unc.  J} & \colhead{H} & \colhead{unc. H} & \colhead{K} & \colhead{unc. K} & \colhead{3.6} & \colhead{unc. 3.6} & \colhead{4.5} & \colhead{unc. 4.5} & \colhead{5.8} & \colhead{unc. 5.8} & \colhead{8.0} & \colhead{unc. 8.0} & \colhead{24} & \colhead{unc. 24} & \colhead{A$_K$} & \colhead{Class$_{Color}$} \\
\colhead{ID}  & \colhead{(J2000)} & \colhead{(J2000)} &\colhead{mag} & \colhead{mag} & \colhead{mag} & \colhead{mag} & \colhead{mag} & \colhead{mag} & \colhead{mag} & \colhead{mag} & \colhead{mag} & \colhead{mag} & \colhead{mag} & \colhead{mag} & \colhead{mag} & \colhead{mag} & \colhead{mag} & \colhead{mag} & \colhead{mag} & \colhead{}  
}

\startdata

1 & 18h29m15.168s & -2d13m09.1286s & 16.09 & 0.11 & 14.43 & 0.05 & 13.48 & 0.04 & 12.67 & 0.01 & 12.21 & 0.01 & 12.01 & 0.02 & 11.60 & 0.03 & 9.03 & 0.08 & 0.931442 & 2 \\
2 & 18h29m26.4168s & -2d12m30.7631s & 16.30 & 0.15 & 14.70 & 0.06 & 13.93 & 0.07 & 13.08 & 0.01 & 12.83 & 0.01 & 12.61 & 0.03 & 12.23 & 0.04 & ... & ... & 1.01858 & 2 \\
3 & 18h29m14.0184s & -2d12m13.329s & 15.58 & 0.07 & 14.05 & 0.03 & 13.19 & 0.03 & 12.41 & 0.01 & 12.06 & 0.01 & 11.81 & 0.02 & 11.55 & 0.03 & 7.15 & 0.02 & 0.803549 & 3 \\
4 & 18h29m30.4608s & -2d11m56.9324s & ... & ... & 15.89 & 0.17 & 14.93 & 0.16 & 13.59 & 0.01 & 13.17 & 0.01 & 12.67 & 0.04 & 11.86 & 0.03 & 8.68 & 0.05 & 0.629381 & 2 \\
5 & 18h29m27.1512s & -2d11m26.8501s & 14.48 & 0.04 & 12.45 & 0.02 & 11.47 & 0.03 & 10.48 & 0.00 & 10.14 & 0.00 & 9.74 & 0.00 & 9.28 & 0.00 & 6.74 & 0.02 & 1.48406 & 2 \\
6 & 18h29m43.9704s & -2d12m55.4202s & ... & ... & ... & ... & ... & ... & 14.20 & 0.03 & 12.61 & 0.01 & 12.26 & 0.04 & 12.28 & 0.08 & 4.00 & 0.00 & ... & 0 \\
7 & 18h30m17.7168s & -2d12m11.646s & 10.47 & 0.02 & 7.69 & 0.05 & 6.03 & 0.02 & 5.21 & 0.00 & 4.73 & 0.00 & 4.09 & 0.00 & 3.96 & 0.00 & 2.69 & 0.00 & 2.05261 & 2 \\
8 & 18h30m18.36s & -2d11m51.4903s & 15.01 & 0.04 & 13.07 & 0.03 & 12.16 & 0.03 & 11.26 & 0.01 & 10.86 & 0.01 & 10.47 & 0.05 & 9.89 & 0.03 & 7.23 & 0.05 & 1.4001 & 2 \\
\enddata

\end{deluxetable}


\begin{figure}
\epsscale{1}
\plotone{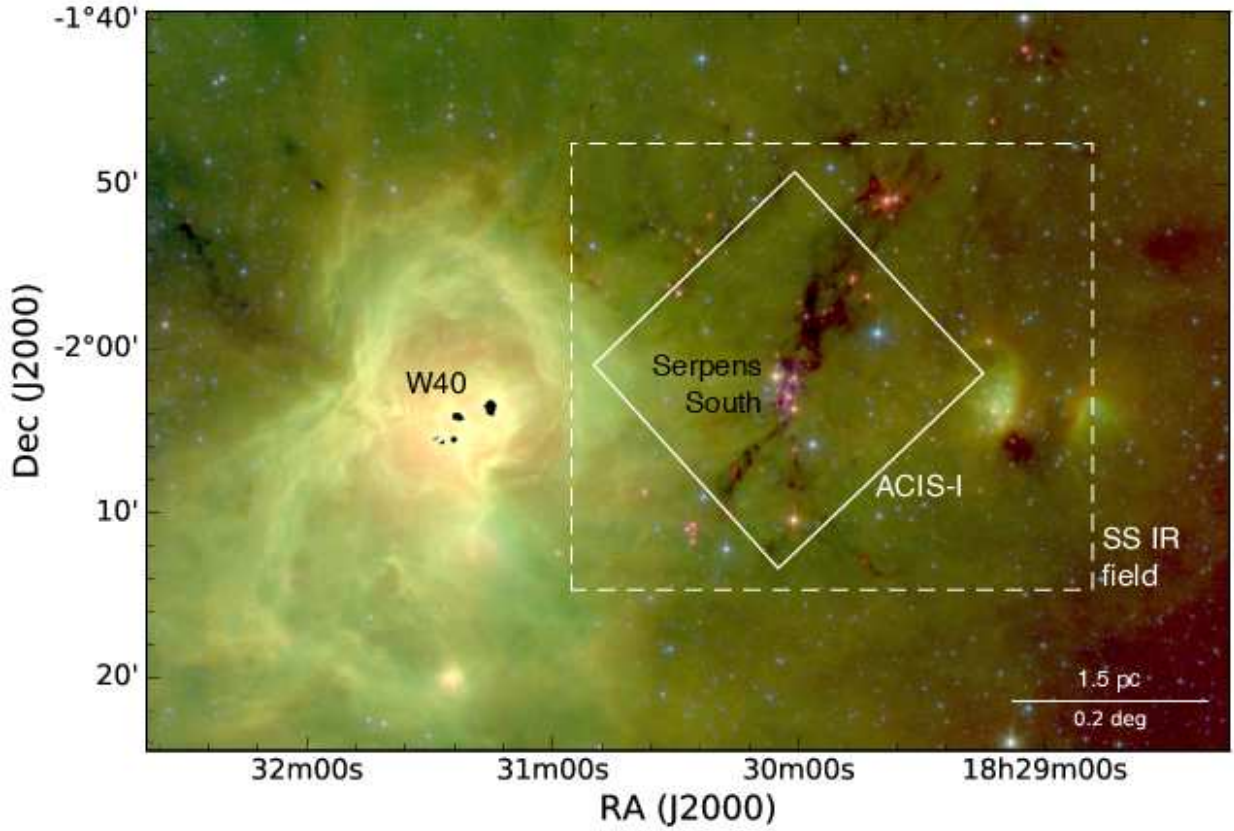}
\caption[]{ 
Spitzer IRAC and MIPS three-colour image of the Serpens South and W40 region, blue: 4.5~${\mu}m$, green: 8.0~${\mu}m$, red: 24~${\mu}m$.
The dark dust filament running through Serpens South is clearly visible, even at 24~${\mu}m$.  
The inner box outlines the Chandra ACIS-I field of view, while the outer dashed box outlines the Serpens South IR field.   
 }
\protect\label{fig1}
\end{figure}

\clearpage

\begin{figure}
\epsscale{1}
\plotone{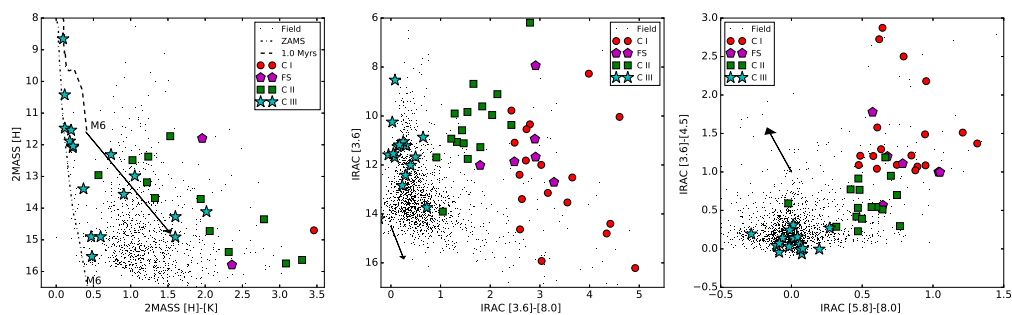}
\caption[]{ 
Selection of color-magnitude and color-color diagrams of the X-ray detected YSOs,  overlaid on all sources in Serpens South field.
{\it Left:} Near-IR 2MASS [H] v [H - K]:  The 1~Myr (dashed) and ZAMS (dash-dot) \citet{sie} isochrones are shown shifted to {\bf260~pc} with no reddening.  
A reddening vector of $2~A_{K}$ is shown.   The {\bf latest} track spectral types plotted are also shown.   
{\it Center:}  IRAC [3.6] v [3.6 - 8.0] diagram: All YSOs show similar range in 3.6$\mu$m mag.  A reddening vector of $5~A_{K}$ is shown.    
{\it Right:} IRAC [3.6 - 4.5] v [5.8 - 8.0]: One Class III and one Class II source show a small excess at longer wavelengths, which might indicate that they have thin outer disks.     A reddening vector of $5~A_{K}$ is shown.    
 
}
\protect\label{fig2}
\end{figure}

\clearpage

\begin{figure}
\epsscale{.80}
\epsscale{1}
\plotone{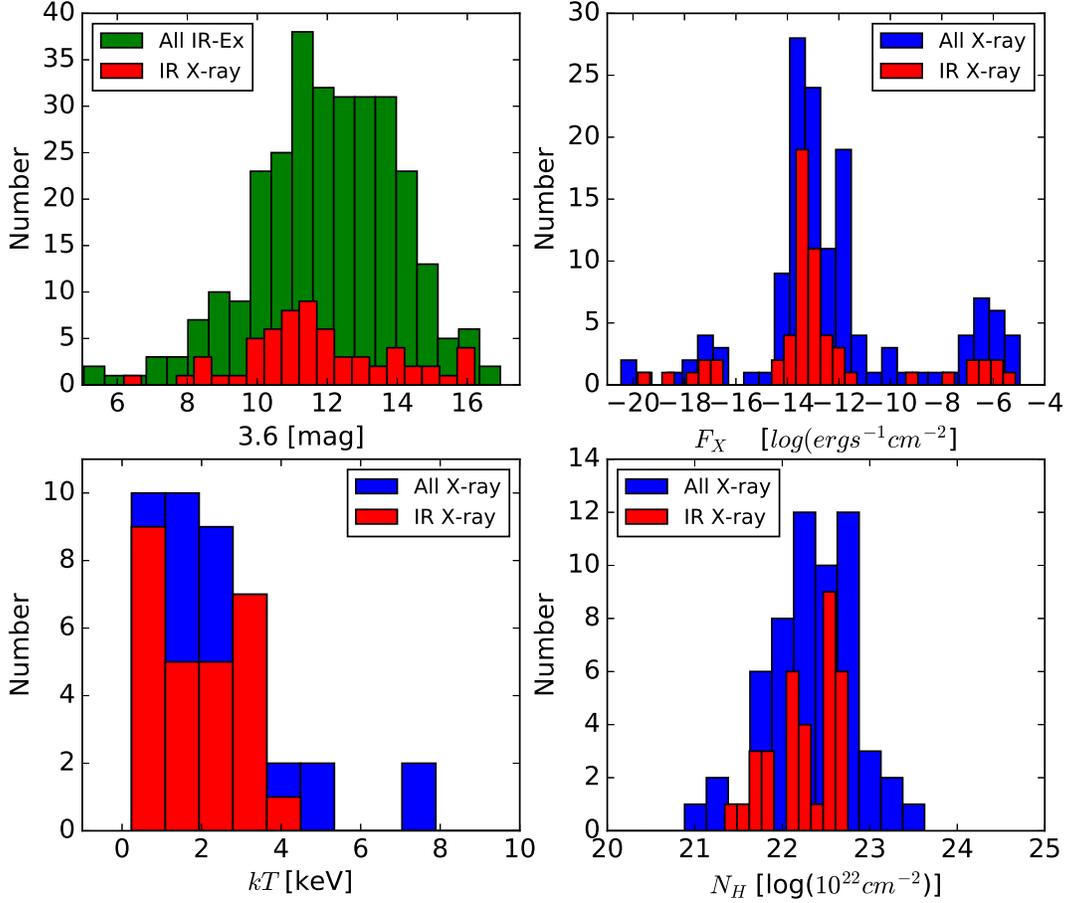}
\caption[]{   
{\it Upper Left:}
Histogram of the IRAC $3.6{\mu}m$ magnitudes for the IR-identified YSOs in the Chandra field (green) and the X-ray identified YSOs (red).  
{\it Upper Right:}
Histogram of X-ray unabsorbed flux, $F_X$, for all X-ray detections (blue) and those with IR counterparts (red). 
{\it Lower Left:}
Histogram of plasma temperature, $kT$, for all X-ray detections (blue) and those with IR counterparts (red). 
{\it Lower Right:}
Histogram of hydrogen column density, $N_H$, for all X-ray detections (blue) and those with IR counterparts (red). 

 }
\protect\label{fig3}
\end{figure}

\clearpage

\begin{figure}
\epsscale{1}
\plotone{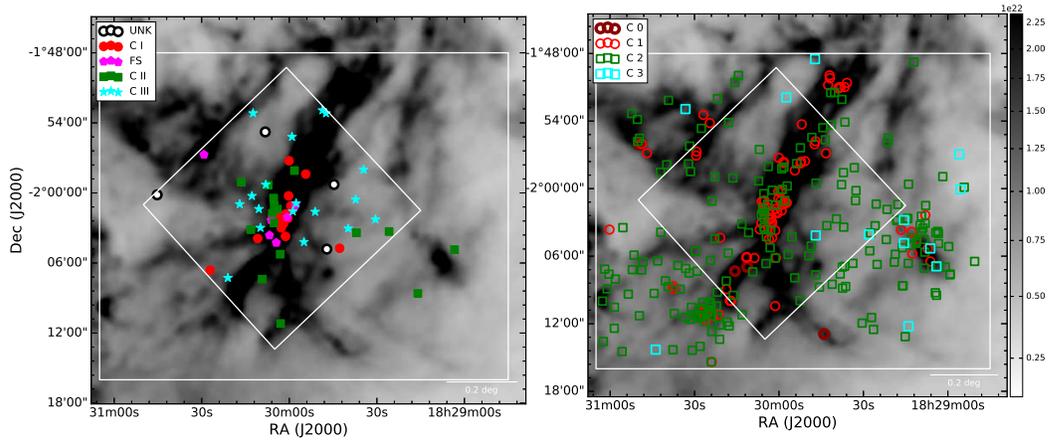}
\caption[]{
Herschel hydrogen column density map showing the distribution of the identified YSOs.   The filled symbols indicate X-ray detected YSOs, the open symbols are detected in the IR only.  
The outline of the Chandra ACIS-I FOV is shown in white, with two detections from the ACIS-S chips also plotted.  
The larger white outline is the SS IR-field.  

}
\protect\label{fig4}
\end{figure}

\clearpage

\begin{figure}
\epsscale{0.75}
\plotone{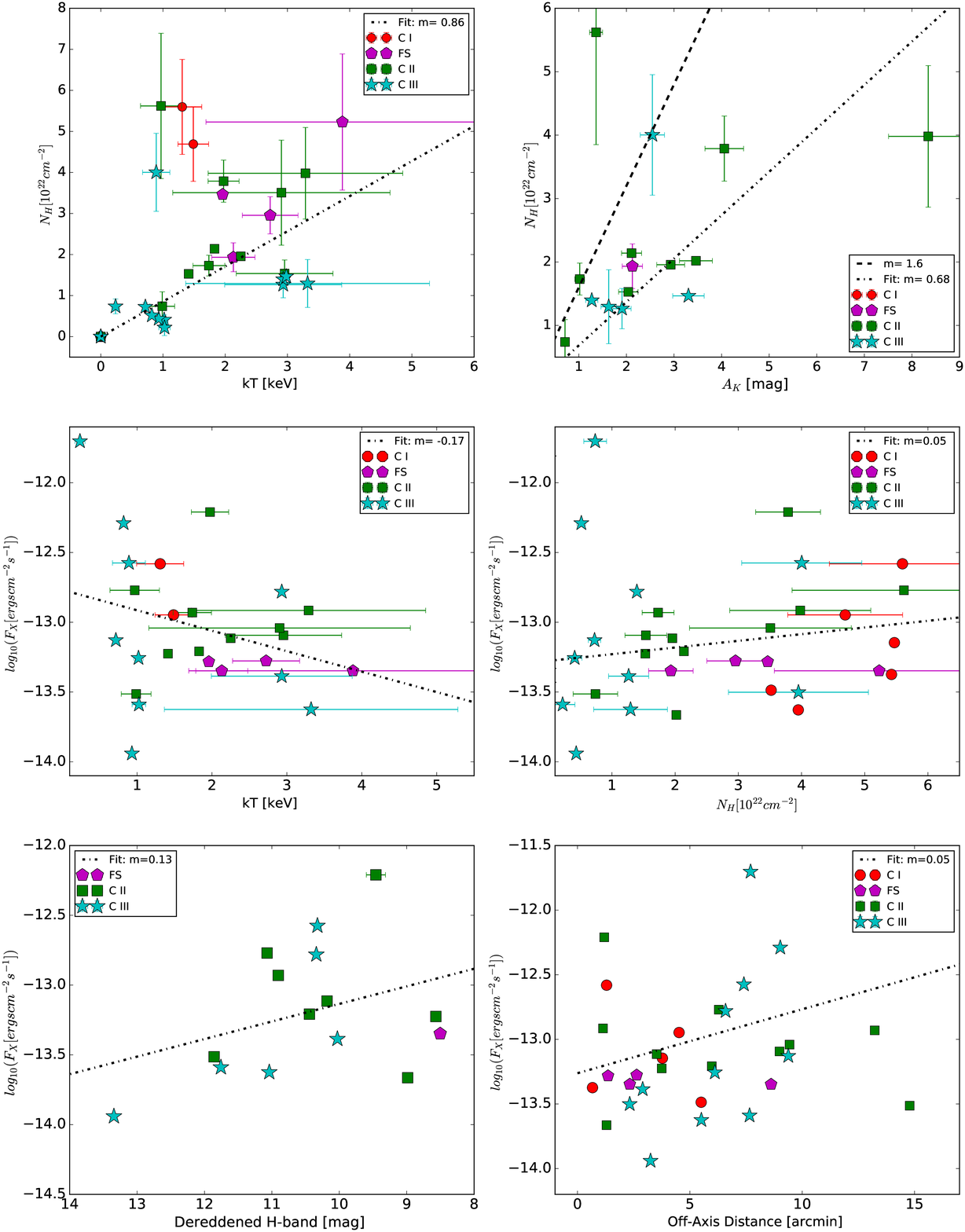}
\caption[]{ 
{\it Upper Left:}
Hydrogen column density, $N_H$, against plasma temperature, $kT$.  
{\it Upper Right:}
Hydrogen column density, $N_H$, against $K$-band extinction, $A_K$.  The dashed line shows the standard gas-to-dust ratio of 1.6, the dashed-dot line shows the best linear regression fit to the data with $N_H = 0.68\times10^{22}  A_K$.  
{\it Center Left:}
X-ray flux, $F_X$, against plasma temperature, $kT$. 
{\it Center Right:}
X-ray flux, $F_X$, against hydrogen column density, $N_H$.  
{\it Lower Left:}
X-ray flux, $F_X$, against dereddened $H$-band magnitude.  
{\it Lower Right:}
X-ray flux, $F_X$, against off-axis angle $\Theta$. 

}
\protect\label{fig5}
\end{figure}

\clearpage

\begin{figure}
\epsscale{0.75}
\plotone{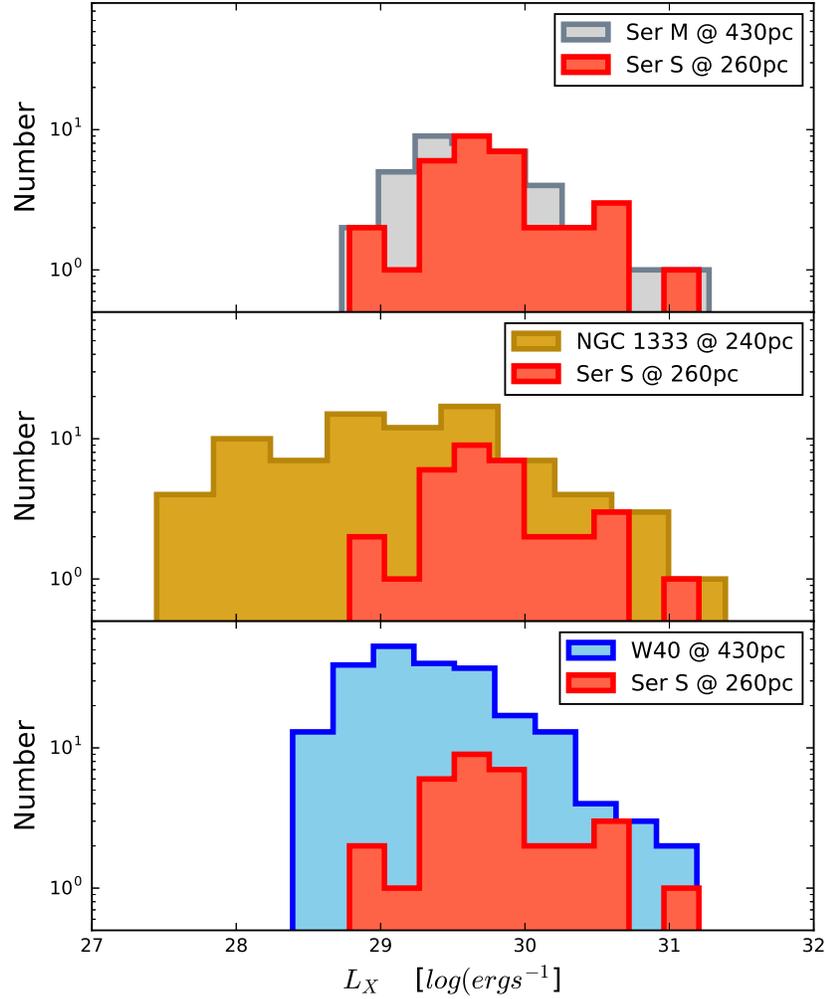}
\caption[]{
Histogram of X-ray luminosity ($L_X$) for Serpens South in comparison with Serpens Main (upper), NGC~1333 (centre), and W40 (lower plot).   The Serpens South distance is taken to be 260~pc.  Serpens Main and W~40 are assumed to lie at a distance of $\sim$430pc \citep{ortiz},  with NGC~1333 at 240pc.   
}
\protect\label{fig6}
\end{figure}

\clearpage

\begin{figure}
\epsscale{0.75}
\plotone{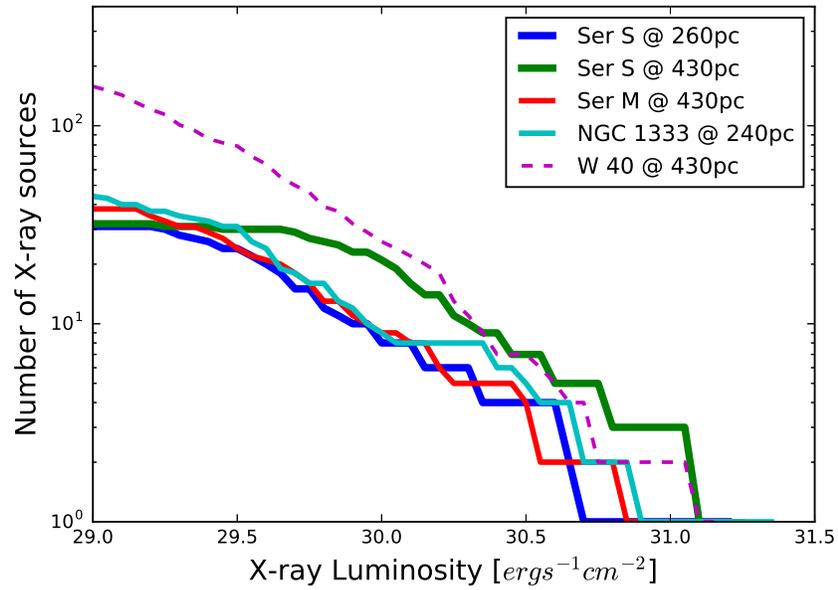}
\caption[]{
 X-ray luminiosity cumulative distributions for Serpens South (at 260 and 430pc), NGC~1333,  Serpens Main, and W~40.   The Serpens South luminosity at 260~pc is consistent with those of Serpens Main at 430~pc and NGC~1333 at 230~pc.   
}
\protect\label{fig7}
\end{figure}

\clearpage

\begin{figure}
\epsscale{1.}
\plotone{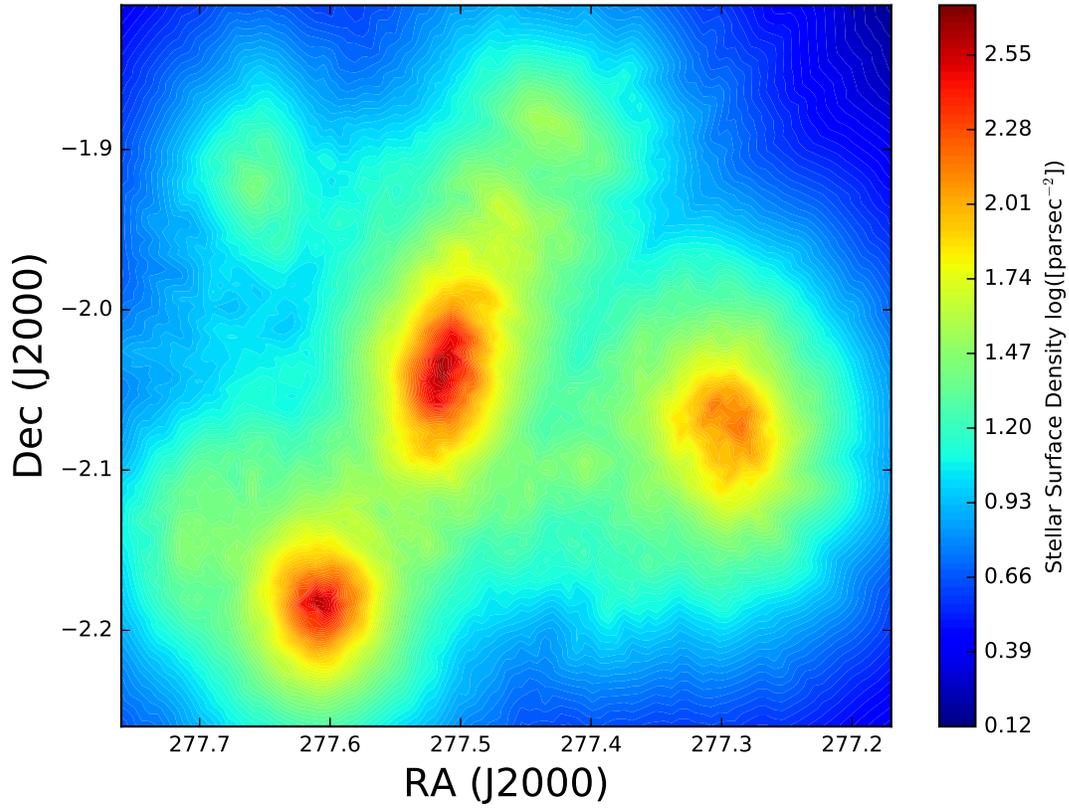}
\caption[]{
Surface Density plot of the identified YSOs in the IR-field, showing four sub regions. The brightest central region is the Serpens South core, 
with a concentration to the SE and an extended ridge to the NW outlining the filamentary structure of the region.  There is a separate cluster 
to the W which may be associated with Serpens South.  
}
\protect\label{fig8}
\end{figure}

\clearpage

\begin{figure}
\epsscale{1.}
\plotone{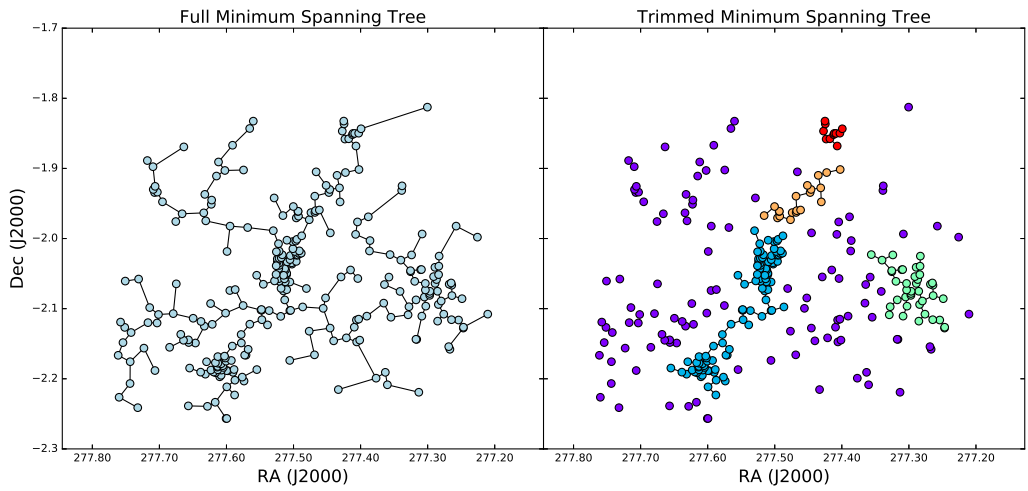}
\caption[]{
MST for the  class 0/1  and class 2 YSOs surrounding Serpens South.  The left subplot shows the full minimum spanning tree.  The right subplot 
shows the trimmed minimum spanning tree with the cutoff set to the characteristic branch length.  
Four clusters are identified in the region:  the Serpens South core, two dense subclusters 
to the north and south following the filament, and an older subcluster to the south-west.   Another possible ring-like young subcluster to the north-east 
does not contain enough members to be selected using the MST method.   
}
\protect\label{fig9}
\end{figure}

\clearpage

\begin{figure}
\epsscale{1.}
\plotone{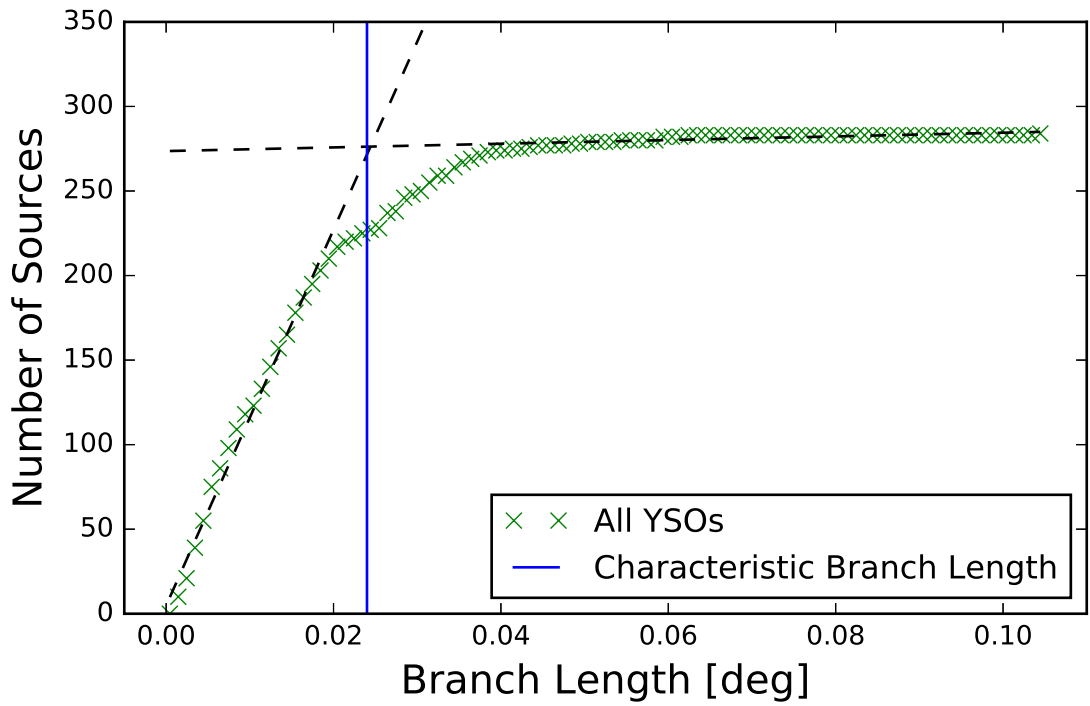}
\caption[]{
MST branch length for all YSOs, and the class I, flat spectrum, and class II sources.  Two linear fits were applied and the 
characteristic branck length was determined to be 0.024 deg or 87'', corresponding to 0.11pc and 0.18pc, at 
cluster distances of 260pc and 430pc, respectively.  
}
\protect\label{fig10}
\end{figure}

\clearpage

\begin{figure}
\epsscale{1.}
\plotone{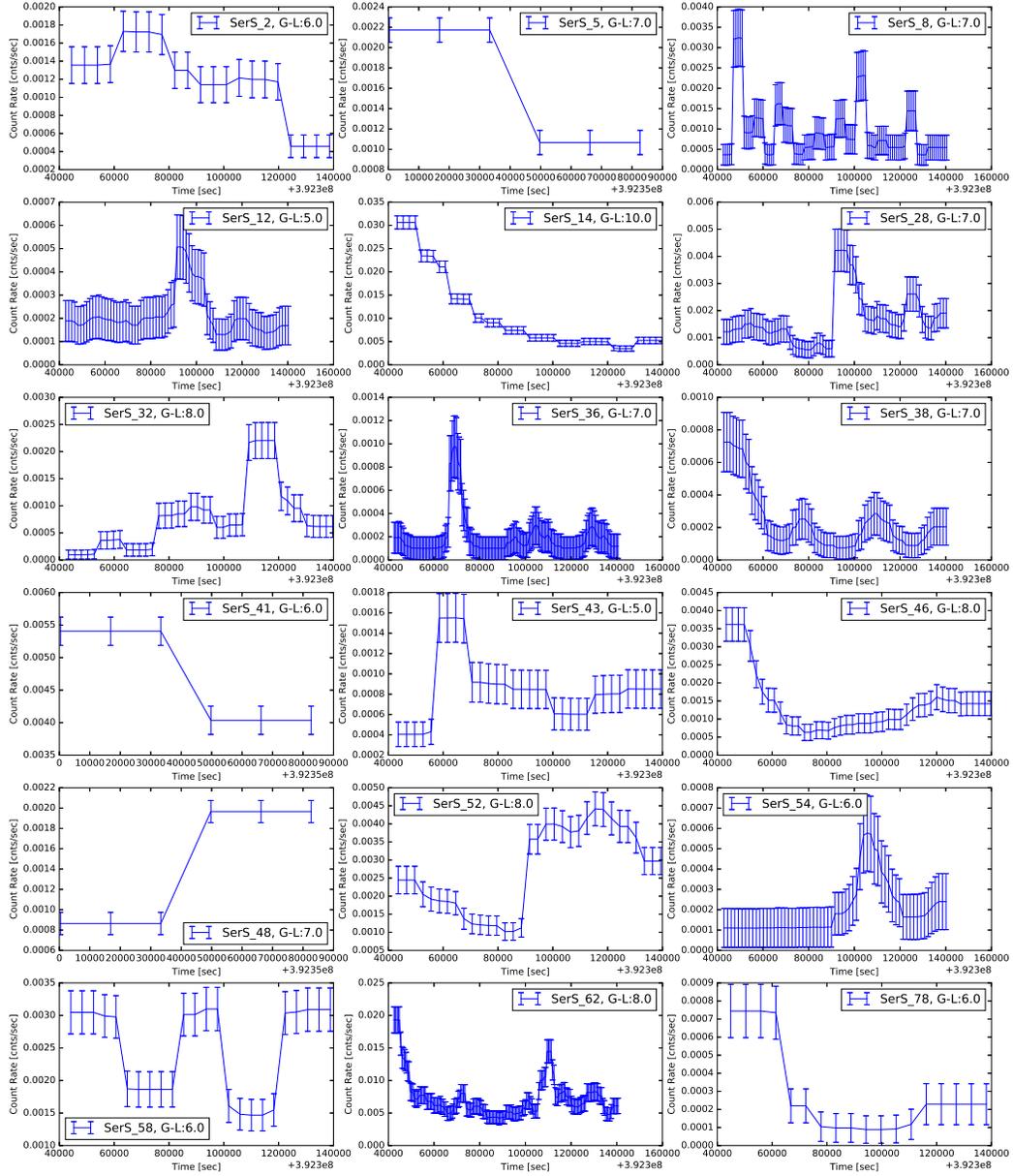}
\caption[]{
Light curves of the 18 identifed variable X-ray sources.  There are 8 flaring sources, and no trend with class in either detection or flaring rate.  
The transition disk source {\it Chandra} ID\#58 shows periodic variability with a period of $\sim$0.46~days.      
}
\protect\label{fig11}
\end{figure}

\end{document}